\documentstyle[aps,preprint,tighten,floats,epsfig,rotating]{revtex}
\begin{document}
\newcommand{\nc}{\newcommand}
\nc{\be}{\begin{equation}}
\nc{\ee}{\end{equation}}
\nc{\bea}{\begin{eqnarray}}
\nc{\eea}{\end{eqnarray}}
\nc{\bib}{\bibitem}
\nc{\1}{\mbox{$16^3\times{32}$}}
\nc{\2}{\mbox{$24^3\times{36}$}}
\nc{\al}{\alpha}
\nc{\alprm}{\al^{\prime}}
\nc{\alprmdlt}{\al^{\prime\delta}}
\nc{\bt}{\beta}
\nc{\bti}{\beta^{\rm{I}}}
\nc{\napefv}{n_{\rm{ape}}(0.50)}
\nc{\niapefv}{n_{\rm{Iape}}(0.50)}
\nc{\napeprm}{n_{\rm{ape}}(\al^{\prime})}
\nc{\niapeprm}{n_{\rm{Iape}}(\al^{\prime})}
\nc{\ncl}{\mbox{$n_{\rm{c}}$}}
\nc{\nicl}{\mbox{$n_{\rm{Ic}}$}}
\nc{\nape}{\mbox{$n_{\rm{ape}}(\alpha)$}}
\nc{\niape}{\mbox{$n_{\rm{Iape}}(\alpha)$}}
\nc{\sutwo}{\mbox{$SU(2)$}}
\nc{\suthree}{\mbox{$SU(3)$}}
\nc{\sun}{\mbox{$SU(N)$}}
\nc{\szero}{\mbox{$S_{0}$}}
\nc{\szr}{\mbox{$\szero=8\pi^2/g^2$}}
\nc{\content}{C_{\rm{L}}=S/\szero}
\nc{\ql}{Q_{\rm{L}}}
\nc{\qil}{Q^{\rm{Imp}}_{\rm{L}}}
\nc{\qimpx}{Q^{\rm{Imp}}_{\rm{L}}(x)}
\nc{\qx}{Q_{\rm{L}}(x)}
\nc{\link}{U_{\mu}(x)}
\nc{\obo}{1\times{1}}
\nc{\oa}{\mbox{${\cal O}(a)$}}
\nc{\oasq}{\mbox{${\cal O}(a^2)$}}
\nc{\hm}{\hat{\mu}}
\nc{\hn}{\hat{\nu}}
\nc{\dg}{\dagger}
\nc{\nm}{\nonumber}
\nc{\realtrace}{{\rm{Re\, Tr}}}
\nc{\trace}{{\rm{Tr}}}
\draft

\preprint{\vbox{\it Submitted to Phys. Rev. D
\hfill\rm ADP-01-23/T457 
}}

\title{\bf Improved Smoothing Algorithms for Lattice Gauge Theory}

\author{Fr\'{e}d\'{e}ric D.R. Bonnet\footnote{
E-mail:~fbonnet@physics.adelaide.edu.au ~$\bullet$~ Tel:
+61~8~8303--3428 ~$\bullet$~ Fax: +61~8~8303--3551},
Derek B. Leinweber,\footnote{E-mail:~dleinweb@physics.adelaide.edu.au
~$\bullet$~ Tel: +61~8~8303--3423 ~$\bullet$~ Fax: +61~8~8303--3551
\hfill\break
\null\quad\quad
WWW:~http://www.physics.adelaide.edu.au/theory/staff/leinweber/}
Anthony G. Williams\footnote{E-mail:~awilliam@physics.adelaide.edu.au
~$\bullet$~ Tel: +61~8~8303--3546 ~$\bullet$~ Fax: +61~8~8303--3551
\hfill\break
\null\quad\quad
WWW:~http://www.physics.adelaide.edu.au/cssm/},
and James M. Zanotti\footnote{E-mail:~jzanotti@physics.adelaide.edu.au
~$\bullet$~ Tel: +61~8~8303--3546 ~$\bullet$~ Fax: +61~8~8303--3551.}.}
\address{Special Research Center for the Subatomic Structure of
Matter (CSSM) and Department of Physics and Mathematical Physics,
University of Adelaide 5005, Australia.} 
\date{\today}
\maketitle

\begin{abstract}

The relative smoothing rates of various gauge field smoothing
algorithms are investigated on ${\cal O}(a^2)$-improved $\suthree$ Yang--Mills
gauge field configurations. In particular, an ${\cal O}(a^2)$-improved version of
APE smearing is motivated by considerations of smeared link projection
and cooling. The extent to which the established benefits of improved
cooling carry over to improved smearing is critically examined. We
consider representative gauge field configurations generated with an
${\cal O}(a^2)$-improved gauge field action on $\1$ lattices at
$\beta=4.38$ and $\2$ lattices at $\beta=5.00$ having lattice spacings
of 0.165(2) fm and 0.077(1) fm respectively. While the merits of improved
algorithms are clearly displayed for the coarse lattice spacing, the
fine lattice results put the various algorithms on a more equal
footing and allow a quantitative calibration of the smoothing rates
for the various algorithms. We find the relative rate of variation in
the action may be succinctly described in terms of simple calibration
formulae which accurately describe the relative smoothness of the gauge
field configurations at a microscopic level.

\end{abstract}

\newpage

\section{Introduction}

Gauge field smoothing algorithms are now widely used in lattice gauge
theory studies as effective tools for constructing operators providing
enhanced overlap between the vacuum and the hadronic state under
investigation. APE smearing~\cite{ape} is now widely used in creating improved
operators for static quark potential studies, or creating orbitally
excited and hybrid mesons from the vacuum. Studies of perfect actions
have lead to the construction of ``Fat Link'' fermion
actions~\cite{DeGrand,DeGrand:1999gp,bernard,jzanottiflic,Melnitchouk,kamleh,kamlehproc}
in which the links appearing in the fermion action are APE smeared. Such
actions display better chiral behavior and reduced exceptional
configuration problems.

Both cooling and smearing algorithms have been used extensively in
studies of QCD vacuum structure, where the lattice operators of
interest suffer from large multiplicative renormalizations
\cite{campostrini,campostrini2}. Here the suppression of short
distance physics is key to removing these perturbative
renormalizations.

Unimproved smoothing algorithms such as standard cooling
\cite{berg,teper1,teper2,Ilgenfritz} or standard APE smearing
\cite{ape} introduce significant errors on each sweep through the
lattice. These errors act to underestimate the action~\cite{perez} and spoil instantons as
the action falls below the one-instanton bound. The problems may be
circumvented by adding additional irrelevant operators to the action
tuned to stabilize instantons
\cite{perez,deforcrand}.

Such improved cooling algorithms are central to studies of topology
and instantons in the QCD vacuum. There thousands of sweeps over the
lattice are required to evolve a typical gauge field configuration to
the self-dual limit. It is well established that the use of improved
algorithms is central to achieving the required level of accuracy.

In this paper we introduce an \oasq -improved form of APE smearing and
examine the extent to which the benefits of improvement in cooling
algorithms carry over to improved smearing. To carefully examine this new
algorithm we create gauge field configurations with an \oasq -improved gauge action.
For this investigation we select Symanzik improvement for the gauge action~\cite{symanzik}.
We consider two sets of gauge field configurations; a coarse \1 lattice at $\beta =
4.38$ with $a \sim 0.165(2)$ fm, and a fine \2 lattice at $\beta = 5.00$
providing $a \sim 0.077(1)$ fm.

These lattices are sufficiently fine that we expect similar results for other
choices of action improvement schemes such as Iwasaki~\cite{iwasaki} or
DBW2~\cite{takaishi}, actions explored in Ref.~\cite{qcd-taro}. We also define an
\oasq -improved non-abelian field strength tensor and construct the corresponding improved
topological charge operator.

While the merits of improved algorithms are clearly displayed for the
coarse lattice spacing, the fine lattice results put the various
algorithms on a more equal footing. Moreover, on the fine lattice we
no longer witness transitions between topological charge values as a
function of smoothing sweeps. 

Finally, we calibrate the relative smoothing rates of standard
cooling, APE smearing, improved cooling and improved smearing
using the action as a measure of the smoothness. The action is selected as it
varies rapidly under cooling. The action evolution has only a mild configuration dependence~\cite{campostrini}
allowing the consideration of only a few configurations in the calibration
process. We focus our computational resources on numerous cooling schemes,
including an improved version of APE smearing referred to as improved smearing.
In all we consider fourteen smoothing algorithms for two hundred sweeps
on seventeen configurations. This is computationally equivalent to the more
standard study of one or two algorithms on the order of a hundred configurations.
This calibration analysis is also an extension of an earlier
analysis~\cite{bonnet} which focused on unimproved algorithms.

The plan of this paper is as follows: Section~\ref{action} describes
the lattice action used in this simulation. The improved field
strength tensor and associated topological charge operator are
described in Section~\ref{toporator}. In motivating improved
smearing we begin in Section~\ref{algos} with a brief review of
improved cooling followed by our improved smearing algorithm.
Section~\ref{numerical} presents the results of our numerical
simulations. The calibration results are discussed in
Section~\ref{calibration} and a summary of the findings is given in
Section~\ref{conclusion}, where the connection to previous
studies drawing relations between cooling and physical
properties~\cite{campostrini,perez2,ringwald} is made.

\section{Lattice Gauge Action}
\label{action}

The tree-level ${\cal O}(a^2)$-improved action is defined as,
\begin{equation}
S_G=\frac{5\beta}{3}\sum_{\scriptstyle x\, \mu\, \nu \atop
\scriptstyle \nu > \mu} 
{\realtrace}(1 - P_{\mu \nu}(x))
-\frac{\beta}{12\, u_{0}^2}\sum_{\scriptstyle x\, \mu\, \nu\, \atop
\scriptstyle \nu > \mu}
{\realtrace}(1 - R_{\mu \nu}(x)),
\label{gaugeaction}
\end{equation}
where $P_{\mu \nu}$ and $R_{\mu \nu}$ are defined as
\begin{eqnarray}
\label{staplessq}
P_{\mu \nu}(x) & = & U_\mu(x)\, U_{\nu}(x+\hm)\, U^{\dg}_{\mu}(x+\hn)
U^\dg_{\nu}(x), \\
\label{staplesrect}
R_{\mu \nu}(x) & = & U_\mu(x)\, U_{\nu}(x+\hm)\, U_{\nu}(x+\hn+\hm)\,
U^{\dg}_{\mu}(x+2\hn)\, U^{\dg}_{\nu}(x+\hn)\, U^\dg_{\nu}(x) \nm \\
& + & U_\mu(x)\, U_{\mu}(x+\hm)\, U_{\nu}(x+2\hm)\,
U^{\dg}_{\mu}(x+\hm+\hn)\, U^{\dg}_{\mu}(x+\hn)\, U^\dg_{\nu}(x).
\end{eqnarray}
The link product $R_{\mu \nu}(x)$ denotes the rectangular $1\times2$
and $2\times1$ plaquettes. $u_0$ is the tadpole improvement factor
that largely corrects for the quantum renormalization of the
coefficient for the rectangles relative to the plaquette.
We employ the plaquette measure for the mean link
\begin{equation}
u_0=\left(\frac{1}{3}{\realtrace}\left< P_{\mu \nu}(x) \right>
\right)^{1/4} \, ,
\label{uzero}
\end{equation}
where the angular brackets indicate averaging over $x$ and $\mu\neq\nu$.

Gauge configurations are generated using the Cabibbo-Marinari~\cite{Cab82}
pseudo-heat-bath algorithm with three diagonal SU(2) subgroups cycled twice.
Simulations are performed using a parallel algorithm on a Thinking Machines
Corporation CM-5 with appropriate link partitioning~\cite{mask}.

Configurations are generated on a $16^3\times{32}$ lattice at
$\beta=4.38$ and a $24^3\times{36}$ lattice at $\beta=5.00$.
Configurations are selected after 5000 thermalization sweeps from a
cold start, and every 500 sweeps thereafter with a fixed mean-link
value. Lattice parameters are summarized in Table~\ref{simultab}.

\begin{table}[b]
\caption{Parameters of the numerical simulations.}
\begin{tabular}{cccccccc}
Action &Volume &$N_{\rm{Therm}}$ & $N_{\rm{Samp}}$ &$\beta$ &$a$ (fm) & $u_{0}$ & Physical Volume fm\\
\hline
Improved & $16^3\times{32}$ & 5000 & 500 & 4.38 & 0.165(2)   & 0.8761 & $2.64^3\times{5.28}$ \\
Improved & $24^3\times{36}$ & 5000 & 500 & 5.00 & 0.077(1)   & 0.9029 & $1.848^3\times{2.772}$ \\
\hline
\end{tabular}
\label{simultab}
\end{table}

\section{Topological Charge Operator}
\label{toporator}

The topological charge of a gauge field configuration provides a
particularly sensitive indicator of the performance of various
smoothing algorithms. The topological charge is related to
the field strength tensor by
\begin{equation}
Q = \sum_{x} q(x) = \sum_{x} \frac{g^2}{32\pi^2} 
{\epsilon_{\mu\nu\rho\sigma}}
{\trace}\left(F_{\mu\nu}(x) F_{\rho\sigma}(x) \right),
\label{latticecharge}
\end{equation}
where $q(x)$ is the topological charge density. An expression for
$F_{\mu \nu}$ may be obtained by expanding the definition of the
Wilson loop. Consider a loop ${\cal C}$ in the $\mu$-$\nu$ plane
\begin{eqnarray}
{\cal{C}}_{\mu\nu}(x) & = & {\cal{P}} \exp \left ( 
i g \oint_{\cal{C}} A(x) \cdot dx \right ) \nm \\
& = & {\cal{P}} \left [ 
1 + i g \left ( \oint_{\cal{C}} A(x) \cdot dx \right) 
  - \frac{g^2}{2!} \left ( \oint_{\cal{C}} A(x) \cdot dx \right )^2
  + {\cal{O}}(g^3) \right ] \, ,
\label{cmunu}
\end{eqnarray}
The line integral is easily evaluated using Stokes theorem and a
Taylor expansion of $\partial_{\mu}A_{\nu}(x)$ about $x_{0}$
\begin{eqnarray}
\oint_{\cal{C}} A(x) \cdot dx  = 
\int dx_{\mu} dx_{\nu} && \biggl [
F_{\mu\nu} (x_{0}) 
+ \left ( x_{\mu}D_{\mu} + x_{\nu}D_{\nu} \right ) F_{\mu\nu}(x_{0})
\nm \\ 
&& \ \left . + \frac{1}{2} \left ( x^{2}_{\mu}D^{2}_{\mu} +
x^{2}_{\nu}D^{2}_{\nu} 
\right ) F_{\mu\nu}(x_{0}) + {\cal{O}}(a^2g^2,a^4) \right] \, .
\label{adotdx}
\end{eqnarray}
The integration limits are determined by the size of the Wilson loop.
Positioning the expansion point $x_{0}$ at the center of the Wilson
loop one finds
\begin{eqnarray}
\label{contour11}
{\oint_{1\times{1}}A(x)\cdot{dx}}&=&a^{2}F_{\mu\nu}(x_{0})+\frac{a^4}{24}\left(D^{2}_{\mu}+D^{2}_{\nu}\right)F_{\mu\nu}(x_{0})+...\\
\label{contour21}
{\oint_{2\times{1}}A(x)\cdot{dx}}&=&2a^{2}F_{\mu\nu}(x_{0})+\frac{a^4}{12}\left(4D^{2}_{\mu}+D^{2}_{\nu}\right)F_{\mu\nu}(x_{0})+...\\
\label{contour12}
{\oint_{1\times{2}}A(x)\cdot{dx}}&=&2a^{2}F_{\mu\nu}(x_{0})+\frac{a^4}{12}\left(D^{2}_{\mu}+4D^{2}_{\nu}\right)F_{\mu\nu}(x_{0})+....
\end{eqnarray}

\begin{figure}[tbp]
\centering{\ 
\epsfig{figure=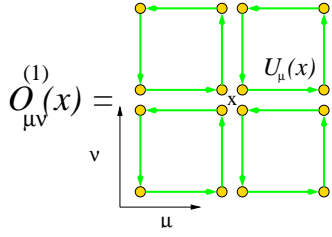,height=3cm} }
\parbox{130mm}{\caption{Graphical representation of the link products
summed in creating ${\cal{O}}^{(1)}_{\mu\nu}(x)$.}
\label{omunu}}
\end{figure}

Hence, $F_{\mu \nu}$ may be extracted from consideration of the $1
\times 1$ plaquette alone. To isolate the second term of the
expansion in Eq.~(\ref{cmunu}), one takes advantage of the Hermitian
nature of $F_{\mu \nu}(x)$. Constructing $F_{\mu \nu}(x)$ symmetrically
about $x$ leads to
\begin{eqnarray}
g\, F_{\mu\nu}(x) & = & \frac{-i}{8} \left [ 
\left ( {\cal{O}}^{(1)}_{\mu\nu}(x)-{\cal{O}}^{(1)\dg}_{\mu\nu}(x)
\right ) 
- \frac{1}{3} Tr \left ( {\cal{O}}^{(1)}_{\mu\nu}(x) -
{\cal{O}}^{(1)\dg}_{\mu\nu}(x) \right) \right] \, ,
\label{fmunu}
\end{eqnarray}
where
${\cal{O}}^{(1)}_{\mu\nu}(x)$ is the sum of
$\obo$ Wilson loops illustrated in Fig.~\ref{omunu},
\begin{eqnarray}
{\cal{O}}^{(1)}_{\mu\nu}(x) &=& U_{\mu}(x)\, U_{\nu}(x+\hm)\,
	     U^{\dg}_{\mu}(x+\hn) \, U^\dg_{\nu}(x) 
             \nm \\
	     &+& U_{\nu}(x)\,
	     U^{\dg}_{\mu}(x+\hn-\hm)\, 
             U^{\dg}_{\nu}(x-\hm)\, U_{\mu}(x-\hm)
	     \nm \\ 
	     &+& U^{\dg}_{\mu}(x-\hm)\,
	     U^{\dg}_{\nu}(x-\hm-\hn)\,
	     U_{\mu}(x-\hm-\hn)\, U_{\nu}(x-\hn)
	     \nm \\ 
	     &+& U^{\dg}_{\nu}(x-\hn)\,
	     U_{\mu}(x-\hn)\, U_{\nu}(x+\hm-\hn)\,
	     U^{\dg}_{\mu}(x) \, .
\label{Omunu}
\end{eqnarray}
This well known definition for $F_{\mu\nu}(x)$ is commonly employed in
the Clover term of the Sheikholeslami-Wohlert~\cite{sheikholeslami}
improved quark action. Although, not everyone enforces the traceless
nature of the Gell-Man matrices by subtracting off the trace as in
Eq.~(\ref{fmunu}). 

Unfortunately, this definition has large \oasq\ errors. These errors
are most apparent in the topological charge. Even after hundreds of
sweeps of cooling this simple definition fails to take on integer
values. Errors are typically at the 10\% level.

\begin{figure}[tbp]
\centering{\
        \epsfig{figure=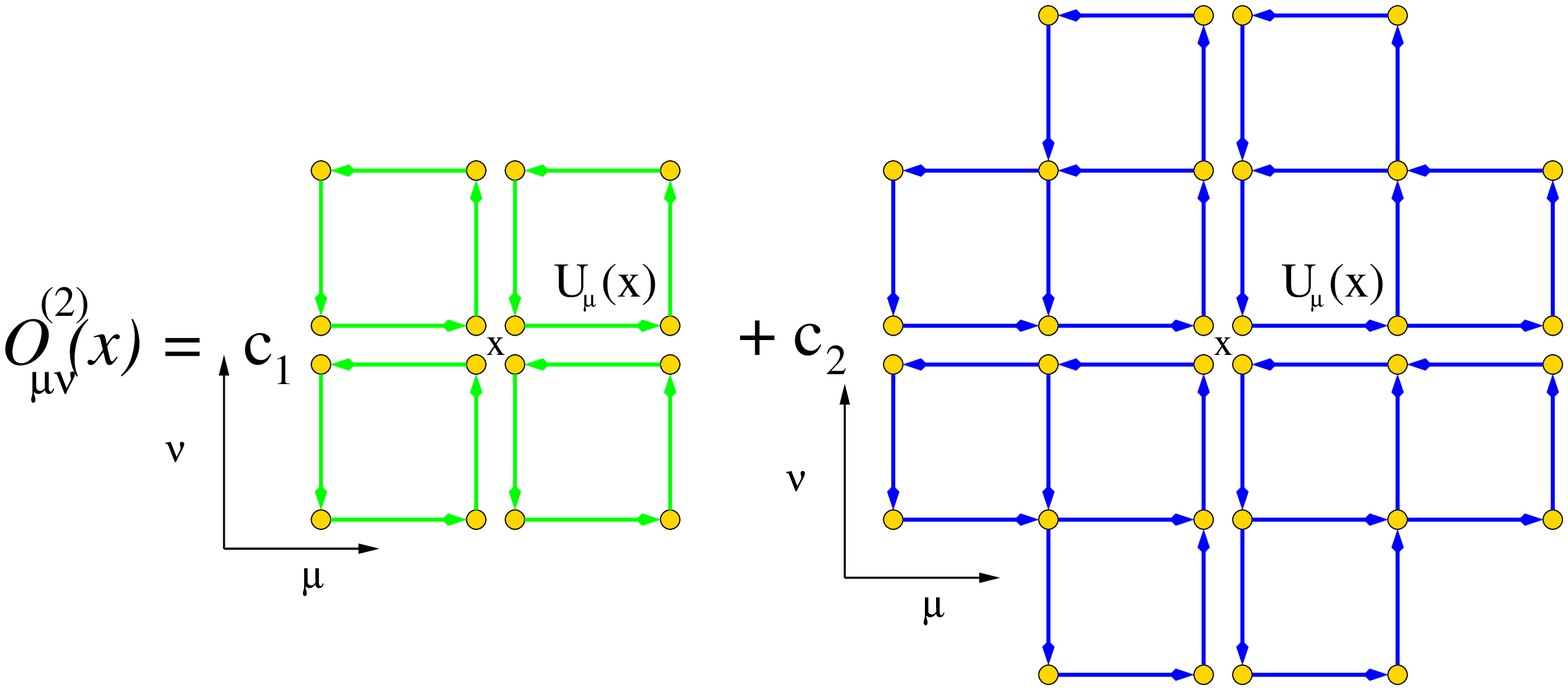,height=5cm} }
\parbox{130mm}{\caption{Graphical representation of the link products
summed in creating ${\cal{O}}^{(2)}_{\mu\nu}(x)$.}
\label{omunu2}}
\end{figure}

To improve the topological charge operator, we improve the definition
of the field strength tensor $F_{\mu \nu}$ by removing \oasq\ errors
with a linear combination of plaquette and rectangle Wilson loops
\begin{equation}
{\cal{O}}^{(2)}_{\mu\nu}(x) = c_{1}{\cal{O}}^{(1)}_{\mu\nu}(x) +
\frac{c_{2}}{u_{0}^{2}}{\cal{I}}^{(2)}_{\mu\nu}(x) \, ,
\label{omunuimp}
\end{equation}
where the tadpole coefficient $u_{0}$ is defined in Eq.~(\ref{uzero}).
Here ${\cal{I}}^{(2)}_{\mu\nu}(x)$ is the link products of $1 \times
2$ and $2 \times 1$ rectangles in the $\mu$-$\nu$ plane
\begin{eqnarray}
{\cal{I}}^{(2)}_{\mu\nu}(x)&=&U_{\mu}(x)U_{\mu}(x+\hm)U_{\nu}(x+2\hm)U^{\dg}_{\mu}(x+\hm+\hn)U^{\dg}_{\mu}(x+\hn)U^\dg_{\nu}(x) \nm \\
&+&U_{\mu}(x)U_{\nu}(x+\hm)U_{\nu}(x+\hm+\hn)U^{\dg}_{\mu}(x+2\hn)U^{\dg}_{\nu}(x+\hn)U^\dg_{\nu}(x)\nm\\
&+&U_{\nu}(x)U_{\nu}(x+\hn)U^{\dg}_{\mu}(x-\hm+2\hn)U^{\dg}_{\nu}(x-\hm+\hn)U^{\dg}_{\nu}(x-\hm)U_{\mu}(x-\hm)\nm\\
&+&U_{\nu}(x)U^{\dg}_{\mu}(x-\hm+\hn)U^{\dg}_{\mu}(x-2\hm+\hn)U^{\dg}_{\nu}(x-2\hm)U_{\mu}(x-2\hm)U_{\mu}(x-\hm)\nm\\
&+&U^{\dg}_{\mu}(x-\hm)U^{\dg}_{\mu}(x-2\hm)U^{\dg}_{\nu}(x-2\hm-\hn)U_{\mu}(x-2\hm-\hn)U_{\mu}(x-\hm-\hn)U_{\nu}(x-\hn)\nm\\
&+&U^{\dg}_{\mu}(x-\hm)U^{\dg}_{\nu}(x-\hm-\hn)U^{\dg}_{\nu}(x-\hm-2\hn)U_{\mu}(x-\hm-2\hn)U_{\nu}(x-2\hn)U_{\nu}(x-\hn)\nm\\
&+&U^{\dg}_{\nu}(x-\hn)U_{\mu}(x-\hn)U_{\mu}(x+\hm-\hn)U_{\nu}(x+2\hm-\hn)U^{\dg}_{\mu}(x+\hm)U^{\dg}_{\mu}(x)\nm\\
&+&U^{\dg}_{\nu}(x-\hn)U^{\dg}_{\nu}(x-2\hn)U_{\mu}(x-2\hn)U_{\nu}(x+\hm-2\hn)U_{\nu}(x+\hm-\hn)U^{\dg}_{\mu}(x)
\, ,
\label{Imunuimp}
\end{eqnarray}
depicted in Fig.~\ref{omunu2}.
The coefficients $c_{1}$ and $c_{2}$ are determined by combining
Eqs.~(\ref{contour11}),~(\ref{contour21}) and~(\ref{contour12}) to
remove \oasq\ errors.
\begin{equation}
a^2\, F_{\mu\nu}(x_{0}) = \frac{5}{3} \left [
{\oint_{1\times{1}}A(x)\cdot{dx}}\right] -
\frac{1}{6}\left[{\oint_{2\times{1}}A(x)\cdot{dx}} + 
{\oint_{1\times{2}}A(x)\cdot{dx}} \right ] \, ,
\end{equation}
indicating the coefficients for the \oasq -improved $F_{\mu \nu}$
of Eq.~(\ref{omunuimp}) are
\begin{eqnarray}
c_{1}=\frac{5}{3}&\hspace*{0.55cm}\rm{and}\hspace*{0.55cm}&c_{2}=-\frac{1}{6}.
\label{topocoef}
\end{eqnarray}

\section{Cooling and Smearing Algorithms}
\label{algos}

In this section we motivate and introduce an improved version of APE smearing. The
motivation is based on improved cooling, and therefore we begin with a
very brief overview of cooling algorithms. Cooling consists of minimizing the
local action effectively one link at the time. The local action is that contribution to the
action associated with a single link.

\subsection{Improved Cooling}
\label{impcool}

Standard cooling minimizes the local Wilson action associated with a
link at each link update. The local action associated with the link
$U_\mu(x)$ is proportional to
\begin{equation}
S_{l}(x,\mu) = \sum_{\scriptstyle \nu \atop \scriptstyle \nu \ne \mu} {\realtrace}
\left ( 1 - U_{\mu}(x) \Sigma_{\mu \nu}(x) \right ) \, ,
\label{localaction}
\end{equation}
where $\Sigma_{\mu \nu}(x)$ is the sum of the two staples associated
with $U_\mu(x)$ lying in the $\mu$-$\nu$ plane
\begin{equation}
\Sigma_{\mu \nu}(x) = U_{\nu}(x+\hm) U^{\dg}_{\mu}(x+\hn) U^\dg_{\nu}(x)
+U^{\dg}_{\nu}(x+\hm-\hn) U^{\dg}_{\mu}(x-\hn) U_{\nu}(x-\hn) \, .
\label{staples}
\end{equation}
The local action associated with the link $\link$ is minimized in standard
cooling by a minimization process, i.e., by replacing the original link by the link
$\link$ which optimizes
\begin{equation}
{\rm{max}}\,{\realtrace} \left ( U_{\mu}(x) \sum_{\scriptstyle \nu \atop \scriptstyle
\nu \ne \mu} \Sigma_{\mu \nu}(x)
\right ) \, . 
\label{maxlocal}
\end{equation}

Improved cooling proceeds in exactly the same manner, but with the
plaquette-based staples replaced with the linear combination of
plaquette-based staples and rectangle-based staples. For improved
cooling we use
\begin{equation}
{\rm{max}}\,{\realtrace} \left ( U_{\mu}(x) \sum_{\scriptstyle \nu \atop \scriptstyle
\nu \ne \mu} \Sigma^I_{\mu \nu}(x),
\right ) \, .
\label{maxlocalimp}
\end{equation}
Where
\begin{equation}
\Sigma^I_{\mu \nu} = {5 \over 3} \, \Sigma_{\mu \nu} + {1 \over 12 \,
u_o^2} \, \Sigma^R_{\mu \nu} \, ,
\label{staplesimp}
\end{equation}
and
\begin{eqnarray}
\Sigma^R_{\mu \nu}(x) & = & U_{\nu}(x+\hm)U_{\nu}(x+\hn+\hm)U^{\dg}_{\mu}(x+2\hn)U^{\dg}_{\nu}(x+\hn)U^\dg_{\nu}(x) \nm \\
& + & U_{\mu}(x+\hm)U_{\nu}(x+2\hm)U^{\dg}_{\mu}(x+\hm+\hn)U^{\dg}_{\mu}(x+\hn)U^\dg_{\nu}(x) \nm \\
& + & U_{\nu}(x+\hm)U^{\dg}_{\mu}(x+\hn)U^{\dg}_{\mu}(x+\hn-\hm)U^{\dg}_{\nu}(x-\hm)U_{\mu}(x-\hm) \nm\\
& + & U^{\dg}_{\nu}(x+\hm-\hn)U^{\dg}_{\nu}(x+\hm-2\hn)U^\dg_{\mu}(x-2\hn)U_{\nu}(x-2\hn)U_{\nu}(x-\hn) \nm\\
& + & U_{\mu}(x+\hm)U^{\dg}_{\nu}(x+2\hm-\hn)U^\dg_{\mu}(x+\hm-\hn)U^\dg_{\mu}(x-\hn)U_{\nu}(x-\hn) \nm\\
& + & U^\dg_{\nu}(x+\hm-\hn)U^{\dg}_{\mu}(x-\hn)U^{\dg}_{\mu}(x-\hn-\hm)U_{\nu}(x-\hn-\hm)U_{\mu}(x-\hm)\, .
\end{eqnarray}
The mean-field factor $u_0$ is updated following each sweep through the
lattice and rapidly goes to 1 as perturbative tadpole contributions
are removed. As we wish to study $\oasq$--improved cooling, the coefficients are based on
those of the action~\cite{symanzik,luscher}. While this action improves
contact with QCD, this choice of action does not completely stabilize instantons
on the lattice~\cite{perez,Iwasaki2}. Removal of $\oasq$ errors and stabilization
of instantons requires the consideration of additional loops~\cite{deforcrand,bilson}.
%
%
The preferred algorithm for finding the $U_\mu(x)$ which maximizes
Eq.~(\ref{maxlocal}) is based on the Cabbibo-Marinari~\cite{Cab82}
pseudo-heat-bath algorithm for constructing \suthree\ -color gauge
configurations. There, operations are performed at the SU(2) level
where the algorithm is transparent.

An element of \sutwo\  may be parameterized as, $U = a_{0}I +
i\,\vec{a}\cdot\vec{\sigma}$, where $a$ is real and $a^{2}=1$. 
Since sums of products of SU(2) matrices are proportional to SU(2)
matrices,
\begin{equation}
\sum_{\scriptstyle \nu \atop \scriptstyle
\nu \ne \mu} \Sigma^I_{\mu \nu}(x) = 
k \overline{U}_\mu(x) \, ,
\end{equation}
where $\overline{U}_\mu(x) \in \sutwo$ and
\begin{equation}
k^2\equiv{\det\left( \sum_{\scriptstyle \nu \atop \scriptstyle
\nu \ne \mu} \Sigma^I_{\mu \nu}(x) \right)} \, . 
\label{coolingk}
\end{equation}
The maximum of the expression
\begin{equation}
{\realtrace} \left ( U_{\mu}(x) \sum_{\scriptstyle \nu \atop \scriptstyle
\nu \ne \mu} \Sigma^I_{\mu \nu}(x)
\right ) = 
{\realtrace} \left ( k\, U_{\mu}(x)\,  \overline{U}_\mu(x) \right ) \, , 
\label{maxlocalimp2}
\end{equation}
is achieved when 
\begin{equation}
{\realtrace} \left ( U_\mu(x) \overline{U}_\mu(x) \right ) = {\realtrace} (I) \, ,
\end{equation}
which requires the link to be updated as
\begin{equation}
U_\mu(x) {\to} U^\prime_\mu(x) = \overline{U}^{-1}_\mu(x) =
\overline{U}^\dagger_\mu(x) =
\left ( \sum_{\scriptstyle \nu \atop \scriptstyle
\nu \ne \mu} {\Sigma^I_{\mu \nu}(x) \over k} \right )^\dagger \, .
\label{uprime}
\end{equation}
At the \suthree\ level, we successively apply this algorithm to the
three diagonal \sutwo\ subgroups of \suthree\ \cite{Cab82}.

When considering an improved cooling algorithm it is crucial to loop
over sufficient SU(2) subgroups to ensure that the subtle effects of
the higher dimension operators introduced in improving the action are
reflected in the final SU(3) link. It is easy to imagine that only
two or three SU(2) subgroups may not take the SU(3) link close enough
to the optimal link for the effects of improvement to be properly
seen. We find consideration of the three diagonal \sutwo\ subgroups
looped over twice to be optimal. In fact we have seen round off
errors actually {\it increase} the action if too many loops are made.

The $u_{0}$ factor of Eq.~(\ref{staplesimp}) is not held fixed during the
improved cooling iteration. Starting from the value determined during
the thermalization process, $u_0$ is updated after every sweep through
the lattice. After a few sweeps of improved cooling the
value quickly converges to 1 which is a good indication that short
distance tadpole effects are being removed in the smoothing
procedure. 

Finally it must be understood that cooling proceeds effectively one
link at a time. That is, links involved in constructing
$\Sigma^I_{\mu \nu}(x)$ are not updated simultaneously with
$U_\mu(x)$. In fact it is a nontrivial task identifying which links
can be simultaneously updated on a massively parallel computer \cite{mask}.

\subsection{Improved Smearing}

In this section we consider the APE smearing~\cite{ape} algorithm. We
extend this algorithm to produce an improved version, motivated by the
success of the improved cooling program.

\subsubsection{Reunitarization of the Links}
\label{runitarization}

APE smearing is a gauge equivariant\footnote{Gauge
equivariance means that if two starting gauge configurations are
related by a gauge transformation then the respective smeared configurations
are also related by the same gauge transformation.}
\cite{hetrick} prescription for smearing a link $\link$ with its
nearest neighbors $U_{\mu}(x+\hn)$, where $\hn$ is transverse to
$\hm$. The APE smearing process takes the form
\begin{eqnarray}
\label{apecomb}
U_{\mu}(x)\longrightarrow{U_{\mu}^\prime(x)} & = & 
(1-\alpha)U_{\mu}(x) + 
\frac{\alpha}{6} \sum_{\scriptstyle \nu \atop \scriptstyle
\nu \ne \mu} \Sigma_{\mu \nu}^\dagger(x) \, , \\
\label{Newreni}
U_{\mu}(x)&=&{\cal P}{U_{\mu}^\prime(x)} \, ,
\end{eqnarray}
where $\Sigma_{\mu \nu}(x)$ is the sum of the two staples associated
with $U_\mu(x)$ defined in Eq.~(\ref{staples}) and ${\cal P}$ is a
projection operator which projects the smeared link back onto
\suthree.

The reunitarization procedure is of central importance when smearing
is applied to the gauge links because the APE smearing operation
produces links outside the \suthree\ gauge group in the intermediate
stage of smearing. The link between smearing and cooling is
established via the manner in which smeared links are projected back
onto SU(3). The preferred approach is to select the link variable
$U_\mu(x)$ that maximizes the quantity
\begin{equation}
{\realtrace} \left ( U_{\mu}(x) U_{\mu}^{\prime\dagger}(x) \right ) \, .
\label{maxAPE}
\end{equation}
In this case a clear connection to cooling is established when $\alpha
\to 1$ as
\begin{equation}
\max {\realtrace} \left ( U_{\mu}(x) U_{\mu}^{\prime\dagger}(x) \right ) 
= \max {\realtrace} \left ( U_{\mu}(x) \sum_{\scriptstyle \nu \atop \scriptstyle
\nu \ne \mu} \Sigma_{\mu \nu}(x) \right ) \, ,
\label{maxAPE1}
\end{equation}
which is precisely the condition of Eq.~(\ref{maxlocal}) for cooling.
In other words, projection of the smeared link back onto the gauge
group via Eq.~(\ref{maxAPE1}) selects the link which minimizes the
local action. Improvement of the staple based on the action will aid
in removing $\oasq$ errors encountered in the $\suthree$ projection.
In practice, the same Cabibbo-Marinari-based cooling method of operating on
$\sutwo$ subgroups may be used to obtain the ultimate SU(3) link.

However there is one very important difference remaining between APE
smearing and cooling. While cooling effectively updates one link at a
time, feeding the smoothed link immediately into the next link update,
APE smearing proceeds uniformly with all links being simultaneously
updated. Smoothed links are not introduced into the algorithm until
the next iteration of the APE smearing process takes place.

It is this latter point that provides the form factor interpretation
of APE smearing in fat-link fermion actions \cite{bernard}. There
smearing can be understood to introduce a form factor suppressing the
coupling of gluons to quarks at the edge of the Brillouin zone where
lattice artifacts are most problematic. The form factor analysis
restricts the smearing fraction to the range $0 \le \alpha \le 3/4$.
Indeed in practice, smearing fractions beyond 3/4 do not lead to
smooth gauge configurations.

\subsubsection{Improving Smearing}
\label{impapesmear}

Having made firm contact with cooling it is clear that replacing the
simple staple of Eq.~(\ref{staples}) with the improved staple of
Eq.~(\ref{staplesimp}) may lead to a realization of the benefits of
improved cooling within APE smearing. Hence we define improved
smearing to be an APE smearing step in which
\begin{eqnarray}
\label{ImpAPEcomb}
U_{\mu}(x)\longrightarrow{U_{\mu}^\prime(x)} & = & 
(1-\alpha)U_{\mu}(x) + 
\frac{\alpha}{6} \sum_{\scriptstyle \nu \atop \scriptstyle
\nu \ne \mu} \Sigma_{\mu \nu}^{I\dagger(x)} \, , \\
\label{NewImpAPE}
U_{\mu}(x)&=&{\cal P}{U_{\mu}^\prime(x)} \, ,
\end{eqnarray}
where $\Sigma_{\mu \nu}^I(x)$ is the improved staple of
Eq.~(\ref{staplesimp}).

Signatures of improvement include the preservation of structures
in the action density distribution under
smearing. \oasq\ errors in the standard Wilson action act to
underestimate the local action and destroy topologically nontrivial
field configurations~\cite{perez}. Improved smearing will have reduced \oasq\
errors and hence better preserve topologically nontrivial field
configurations. Hence an associated signature of improvement is the
stability of topological charge under hundreds of smearing sweeps.
Of course, one could alter the coefficients of the improvement
terms to stabilize instantons~\cite{perez} at the expense of
introducing $\oasq$ errors into the smoothing action.

The extended nature of the staple will alter the stability range of
$\alpha$. Where standard APE smearing provides the range $0 \le
\alpha \le 3/4$, we find the upper stability limit lies below $\alpha
= 0.6$. 

\section{Numerical Simulations}
\label{numerical}

We analyze two sets of gauge field configurations generated using the
Cabibbo-Marinari~\cite{Cab82} pseudo-heat-bath algorithm with three
diagonal \sutwo\ subgroups looped over twice. Details may be found in
Table~\ref{simultab}. As discussed in the introduction, analysis of a few
configurations proves to be sufficient to resolve the nature of the algorithms under
investigation. We consider eleven $\1$ configurations and
six $\2$ configurations. For each configuration we
separately perform 200 sweeps of cooling and 200 sweeps of improved
cooling. We explore 200 sweeps of APE smearing at seven values of the
smearing fraction and 200 sweeps of improved smearing at five values of
the smearing fraction. For APE smearing we consider $\alpha = 0.10$,
0.20, 0.30, 0.40, 0.50, 0.60, and 0.70. Similarly for improved smearing
we consider $\alpha = 0.10$ to 0.50 at intervals of 0.10. The extended
nature of the staple alters the stability range of $\alpha$ to
lie below $\alpha = 0.6$.

For clarity, we define the number of times an algorithm is applied to
the entire lattice as $\ncl$, $\nicl$, ${\nape}$ and $\niape$ for
cooling, improved cooling, APE smearing and improved smearing
respectively. We monitor both the total action normalized to the
single instanton action $S_0 = 8 \pi^2 / g^2$ and the topological
charge operators, $\ql$ and $\qil$, from which we observe their
evolution as a function of the appropriate sweep variable and smearing
fraction $\alpha$.

\subsection{The Influence Of The Number Of Subgroups On The Gauge Group.}
\label{subgroups}

In this section we describe the influence of including additional
\sutwo\ subgroups in constructing the gauge group \suthree\ and
explore the impact it has on the smoothing procedure. The Cabibbo-Marinari
algorithm~\cite{Cab82} constructs the $\sun$
gauge groups using \sutwo\ subgroups. It is understood that the
minimal set required to construct \suthree\ matrices is two diagonal
\sutwo\ subgroups. After having performed cooling on gauge field configurations we
noticed that the resulting cooled gauge field configurations were not
smooth even after a large number of smoothing steps. Adding a third
\sutwo\ subgroup made a significant difference in the resulting
smoothness.

We explored further by simply performing additional cycles,
$n_{\rm{cycle}}$, around the three diagonal \sutwo\  subgroups. We
monitored the smoothing rate using both the Standard Wilson and
the improved action according to the cycle number being set to
$n_{\rm{cycle}}=1,2$ and 3. Based on the evolution of the action, we
found that the optimum cooling rate on gauge field configurations is
achieved using the three diagonal \sutwo\  subgroups cycled over
twice, $n_{\rm{cycle}}=2$. Cycling more than twice provides very little
further reduction of the action and round off errors may actually
increase the action on occasion. Hence two cycles over the three
diagonal \sutwo\ subgroups is sufficient to precisely create the
\suthree\ link which minimizes the local action. This determination
is crucial to ensuring the effects of our improved action are fully
reflected in the \suthree\ link.

\begin{figure}
\centering{\
\epsfig{angle=90,figure=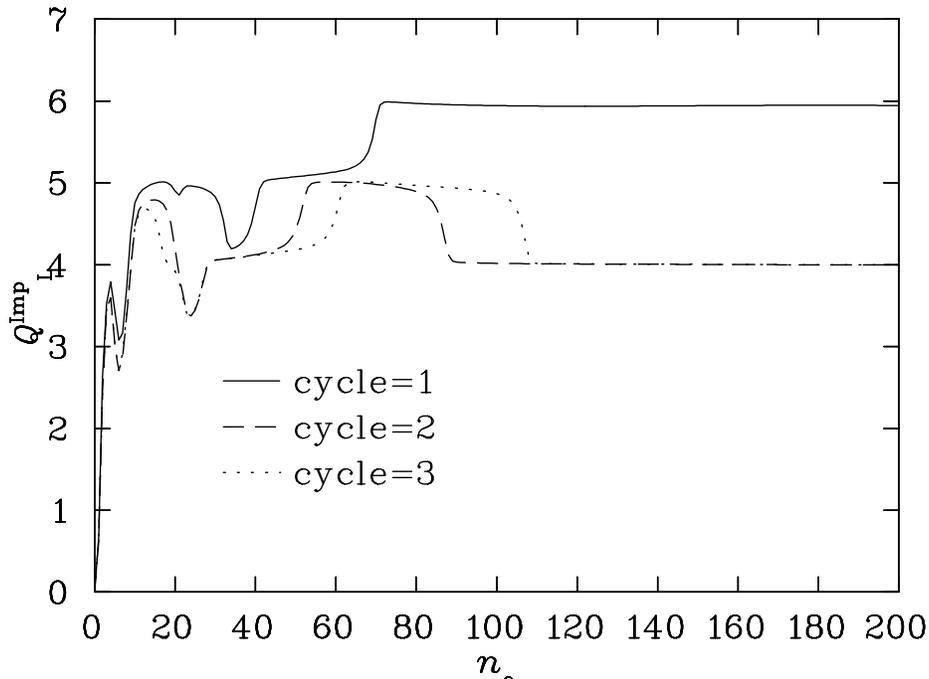,height=9cm} }
\parbox{130mm}{\caption{The evolution of the topological charge estimated by the improved operator as a function of standard cooling sweeps $\ncl$ for various numbers of \sutwo\ subgroups. The curves are for a typical configuration from the $\1$ lattices where $a=0.165(2)$ fm. The parameter cycle describes the number of times the three diagonal \sutwo\ subgroups are cycled over.}
\label{nsubstd}}
\end{figure}

\begin{figure}
\centering{\
\epsfig{angle=90,figure=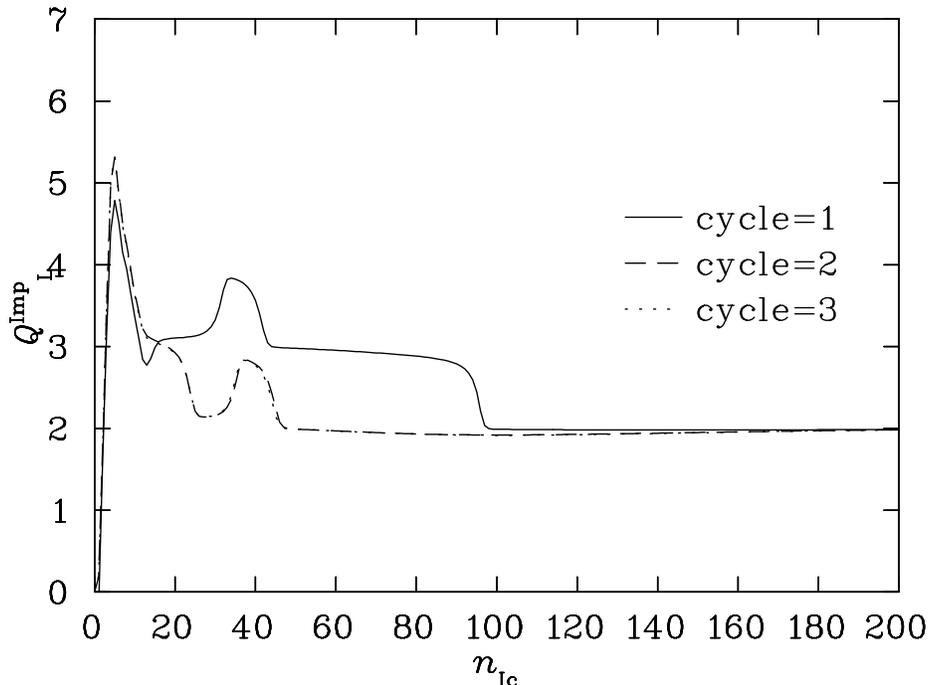,height=9cm} }
\parbox{130mm}{\caption{The evolution of the topological charge estimated by the improved operator as a function of improved cooling sweeps $\nicl$ for various numbers of cycles over the three diagonal \sutwo\ subgroups. The curves are for the same configuration illustrated in Fig.~ \protect\ref{nsubstd}.}
\label{nsubimp}}
\end{figure}

We also monitored the evolution of the topological charge with respect
to the above number of cycles. On the $\1$ lattice with spacing of
$a=0.165(2)$ fm, we observe a disagreement of the trajectories for the
topological charge for different numbers of the \sutwo\ subgroups
cycles. Fig.~\ref{nsubstd} displays results for standard cooling
and Fig.~\ref{nsubimp}
displays similar results for improved cooling. A comparison of Figs.~\ref{nsubstd} and~\ref{nsubimp}
indicates the trajectories also differ between cooling and improved
cooling.

Hence we see subtle differences in the algorithms leading to dramatic
differences in the topological charge. One must conclude that a
lattice spacing of 0.165(2) fm is too coarse for a serious study of
topology in \suthree\ gauge fields using the algorithms considered
here. The difficulty lies in the fact that the algorithms have a dislocation
threshold typically a little over two lattice spacings~\cite{deforcrand,TDeGrand}. Instantons
with a size smaller than the dislocation threshold are removed during the process
of smoothing. While this threshold has the desireable property of removing
lattice artifacts, we note that twice the lattice spacing is 0.33 fm. Minor
differences in the dislocation thresholds of the various algorithms will cause
some (anti)instantons to survive under one algorithm where they are removed under
another.

On the other hand, a lattice spacing the order of 0.077(1) fm appears to allow a
meaningful study of topology in \suthree\ gauge theory. Twice this lattice spacing
is 0.15 fm, well below the typical size of instantons~\cite{ringwald}.

We also note here the accuracy with which our improved topological
charge operator reproduces integer values. These results should be
contrasted with the usual 10\% errors of the unimproved operator at
similar lattice spacings. Such errors on a topological charge of 5
can lead to the uncomfortable result of $Q \simeq 4.5$ when the
unimproved operator is used.

\begin{figure}
\centering{\epsfig{angle=90,figure=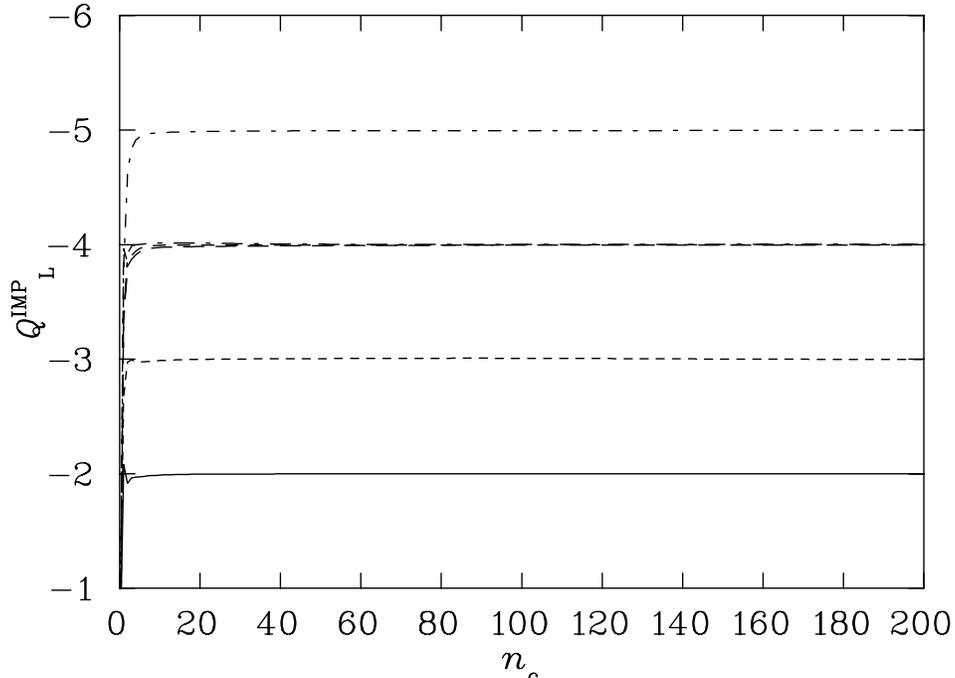,height=9cm} }
\parbox{130mm}{\caption{The evolution curve for the topological charge estimated via the improved operator as a function of cooling sweeps $\ncl$ for six configurations on the $\2$ lattices at $\bt=5.00$ where $a={0.077(1)}$ fm.}
\label{coolIQc000to005}}
\end{figure}

\begin{figure}
\centering{\epsfig{angle=90,figure=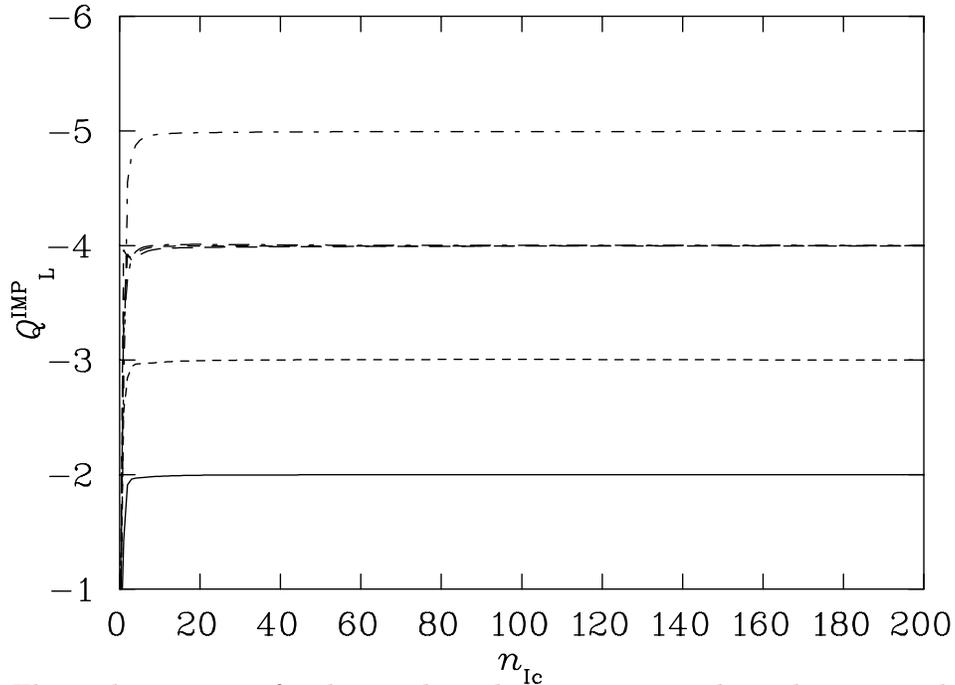,height=9cm} }
\parbox{130mm}{\caption{The evolution curve for the topological charge estimated via the improved operator as a function of improved cooling sweeps $\nicl$ for the same six configurations from the $\2$ lattices at $\bt=5.00$ illustrated in Fig.~\protect\ref{coolIQc000to005}.}
\label{ImpcoolIQc000to005}}
\end{figure}

With the finer $\2$ lattice at $\bt=5.00$, we observe perfect
agreement among trajectories for different numbers of cycles
of the three \sutwo\ subgroups. Moreover, the topological charge
remains stable for hundreds of sweeps following the first three
sweeps. Figures~\ref{coolIQc000to005} and~\ref{ImpcoolIQc000to005}
compare the topological charge evolution for cooling versus improved
cooling for six configurations. In every case, the two algorithms
produce the same topological charge for a given configuration.

\subsection{The Action}
\label{argumentaction}

We begin by considering the action evolution on both lattices. Here
we report the action divided by the single instanton action $\szr$.
It is important to note that although the $\2$ lattice has almost four
times more lattice sites than the $\1$ lattice, the physical volume is
smaller by almost a factor of three. As such the typical topological
charges encountered are smaller in magnitude.

\begin{figure}[tbp]
\centering{\epsfig{angle=90,figure=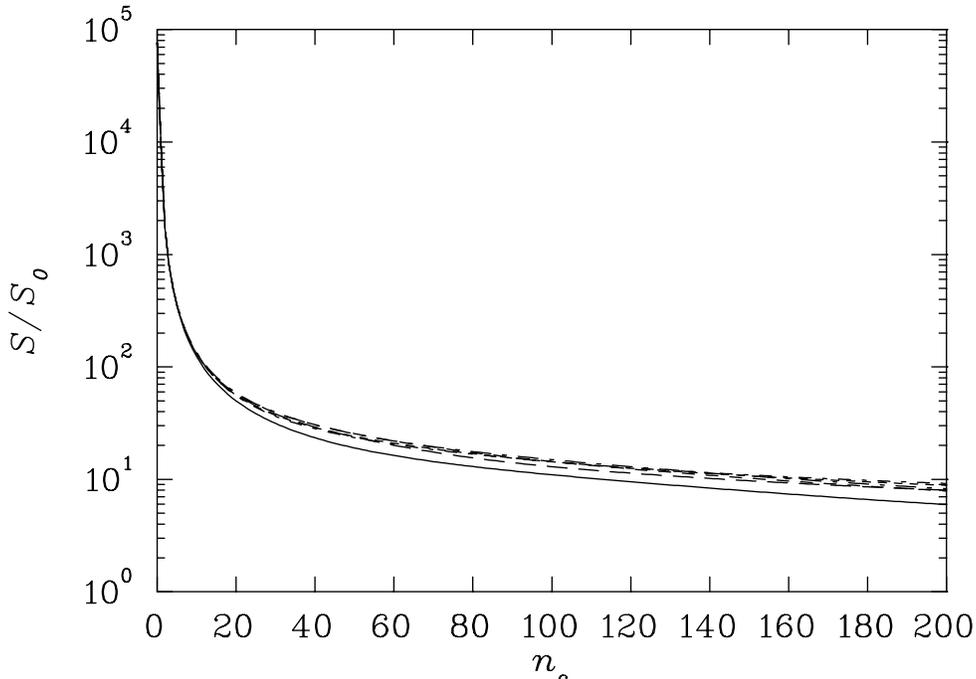,height=9cm} }
\parbox{130mm}{\caption{The ratio $S/S_{0}$ as a function of standard cooling
sweeps $n_{\rm{c}}$ for five configurations on the $\2$ lattice at
$\bt=5.0$. The single instanton action is $S_{0}=8\pi^2/g^2$.}
\label{coolingcurve}}
\end{figure}

\begin{figure}[tbp]
\centering{\epsfig{angle=90,figure=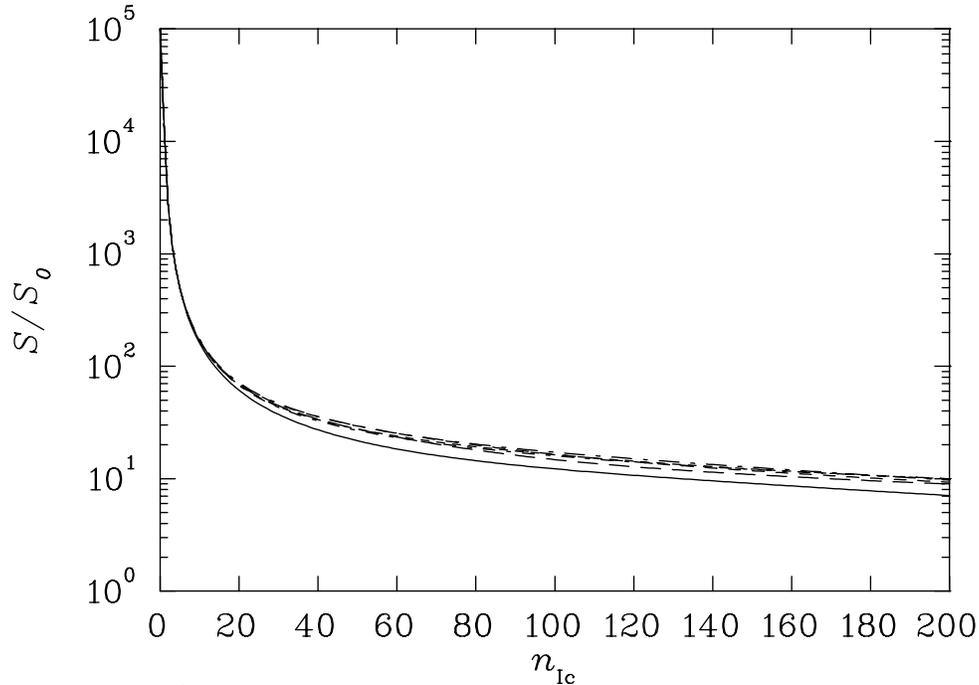,height=9cm} }
\parbox{130mm}{\caption{The ratio $S/S_{0}$ as a function of improved
cooling sweeps $n_{\rm{Ic}}$ for five configurations on the $\2$ lattice at $\bt=5.0$. The rate of cooling is seen to be somewhat slower than that for the standard cooling.}
\label{Impcoolingcurve}}
\end{figure}

\begin{figure}[tbp]
\centering{\epsfig{angle=90,figure=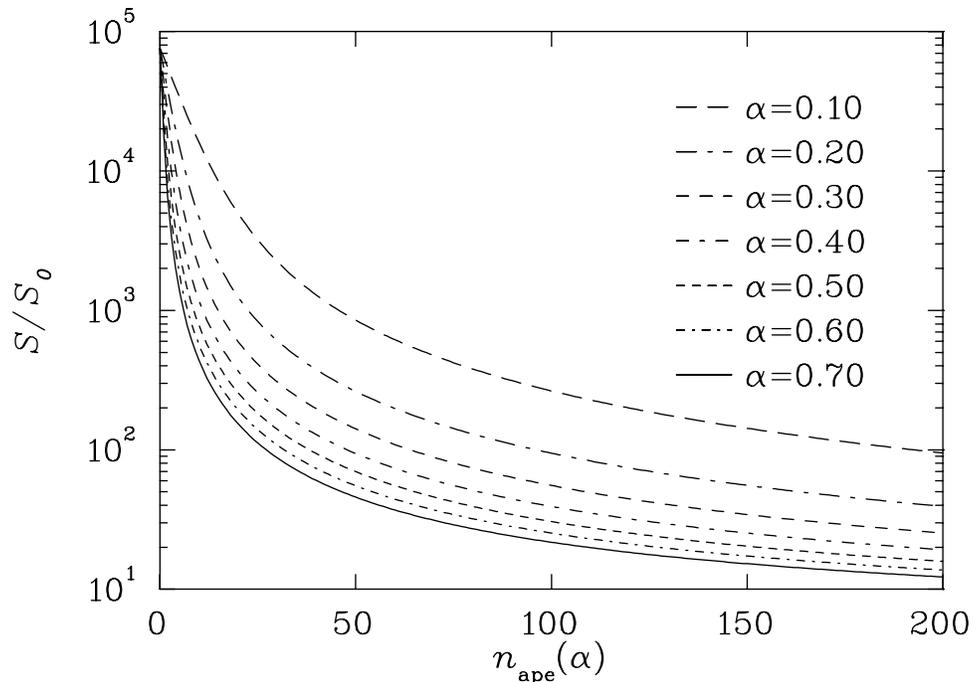,height=9cm} }
\parbox{130mm}{\caption{The ratio $S/S_{0}$ as a function of APE
smearing sweeps $\nape$ for one configuration on the $\2$ lattice at
$\bt=5.0$. Each curve has an associated smearing fraction $\al$. The rate of lowering the action for the maximum stable smearing fraction ($\approx{0.75}$) is seen to be less than that for the other standard or improved cooling.} 
\label{smearingcurve}}
\end{figure}

\begin{figure}[tbp]
\centering{\epsfig{angle=90,figure=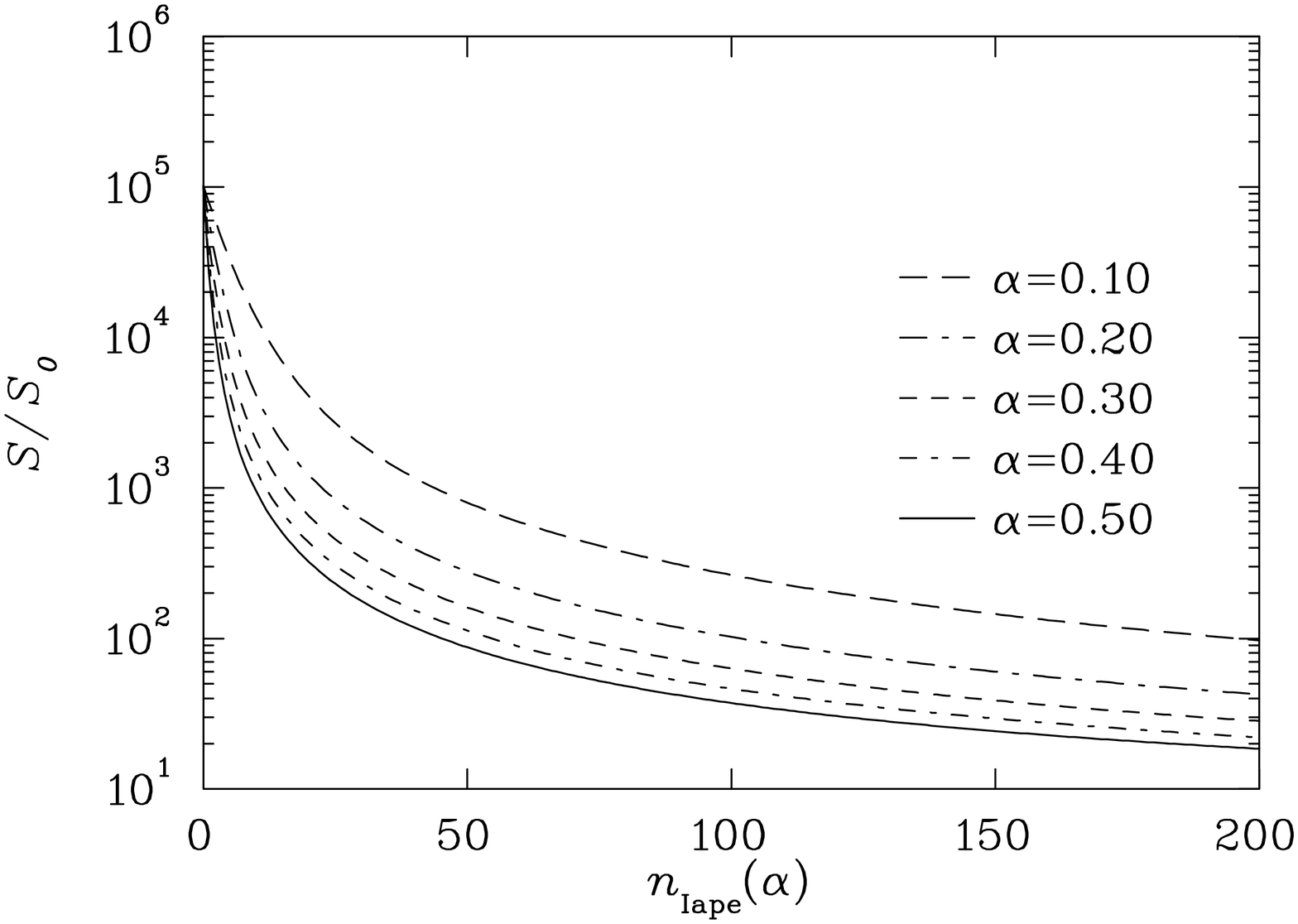,height=9cm} }
\parbox{130mm}{\caption{The ratio $S/S_{0}$ as a function of improved
smearing sweeps $\niape$ for one configuration on the $\2$ lattice at
$\bt=5.0$. Each curve has an associated smearing fraction $\al$. We see that this is the slowest of the four algorithms for lowering the action as a function of the sweep number.}
\label{Impsmearingcurve}}
\end{figure}

Figs.~\ref{coolingcurve},~\ref{Impcoolingcurve},~\ref{smearingcurve},
and~\ref{Impsmearingcurve} report the typical evolution of the action
under standard cooling, improved cooling, APE smearing and improved
smearing respectively. Inspection of the figures reveals that
improved cooling preserves the action better than standard cooling
over a couple hundred sweeps. As expected, standard APE smearing
remains slower than cooling or improved cooling even at our most
efficient smearing fraction ($\al=0.70$). Similar results are
observed for improved smearing at our most efficient smearing fraction
of $\al=0.50$ in Fig.~\ref{Impsmearingcurve}.

Based on these observations, one concludes that the fastest way to
remove the short range quantum fluctuations on an $\oasq$ gauge field
configuration, is through standard cooling, which lowers the action more
rapidly than improved cooling as a function of cooling sweep. In turn
we see that improved cooling is faster than the maximum stable
standard APE smearing, which is faster than the maximum stable improved
smearing. This is illustrated by Figs.~\ref{coolingcurve}--\ref{Impsmearingcurve}.
It is important to emphasize that the fastest way of removing these
fluctuations is not necessarily the best as far as the topology is
concerned. It is already established that the \oasq\ errors of the
standard Wilson action act to underestimate the action~\cite{perez}. These errors
spoil instantons which might otherwise survive under improved cooling.

\subsection{Topological Charge from Cooling and Smearing}
\label{argumenttopo}

\begin{figure}[tbp]
\centering{\epsfig{angle=90,figure=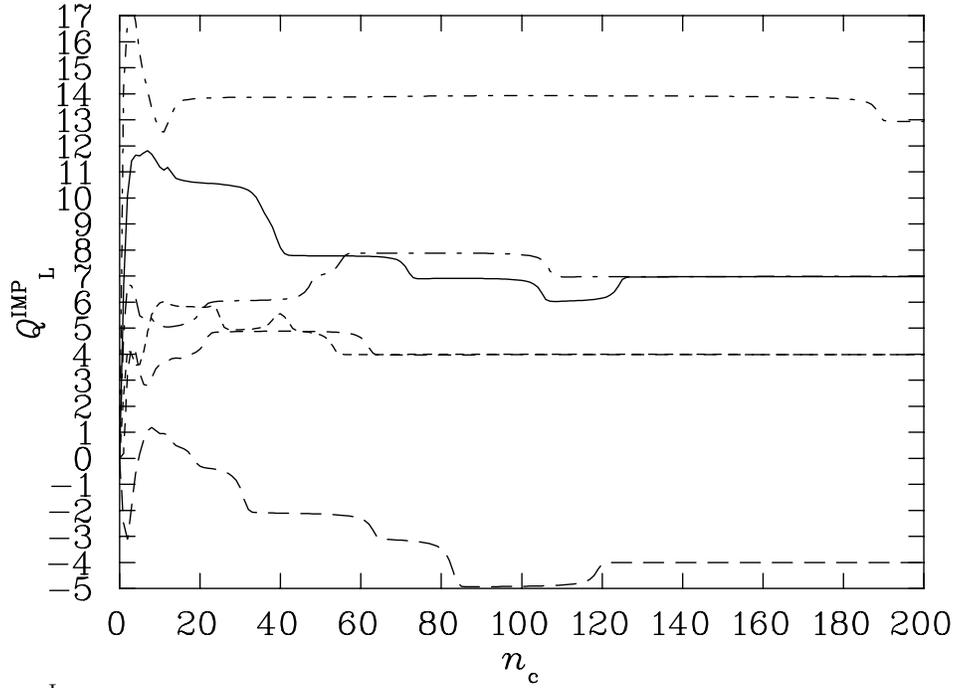,height=9cm} }
\parbox{130mm}{\caption{$\qil$ versus $\ncl$ for six configurations on
the $\1$ lattices at $\bt=4.38$, $a=0.165(2)$ fm. Each line corresponds
to a different configuration. }
\label{coolIQc90}}
\end{figure}

\begin{figure}[tbp]
\centering{\epsfig{angle=90,figure=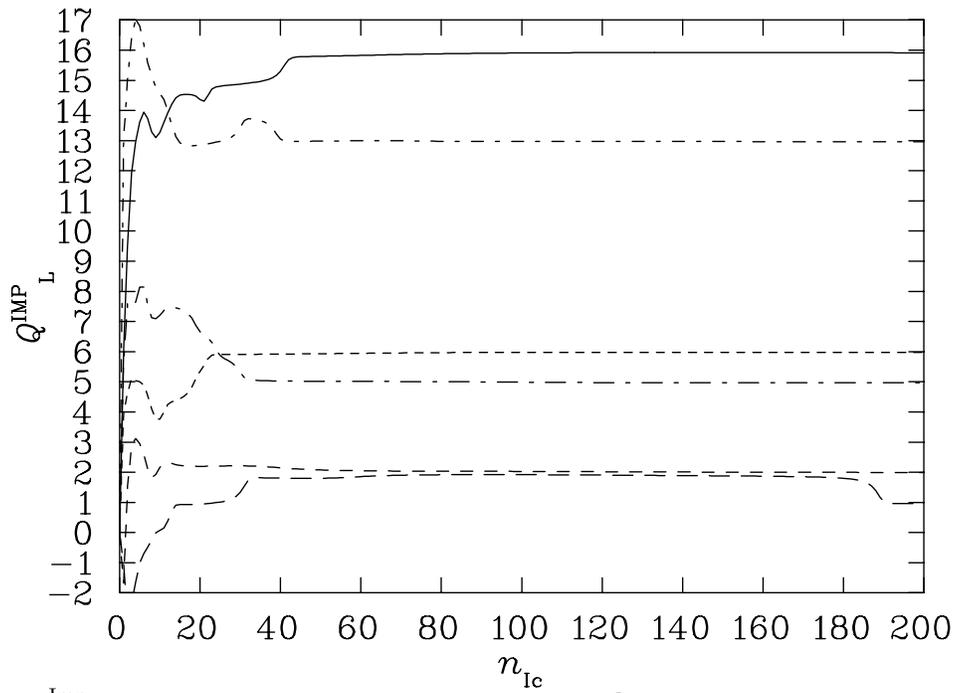,height=9cm} }
\parbox{130mm}{\caption{$\qil$ versus $\nicl$ for six configurations
on the $\1$ lattices. The different line types identifying different
configurations match the configurations identified in Fig.~\protect\ref{coolIQc90}.}
\label{ImpcoolIQc90}}
\end{figure}

We begin by considering the $\1$ lattices having a lattice spacing of
$a=0.165(2)$ fm. In Fig.~\ref{coolIQc90}, we plot the evolution curve
for the improved topological charge as a function of the cooling sweep
number, \ncl, for six of our configurations. Similarly in Fig.~\ref{ImpcoolIQc90},
for the same six configurations we plot the
improved topological charge operator but this time as a function of
improved cooling sweep $\nicl$. The line types in Figs.~\ref{coolIQc90}
and~\ref{ImpcoolIQc90} correspond to the same
underlying configurations and are to be directly compared. For
example, the solid curve corresponds to the same gauge field
configuration in both figures but with a different algorithm applied
to it. From these two figures we notice the two cooling methods lead
to completely different values for the topological charge. 

Improved cooling brings stability to the evolution of the topological
charge whereas standard cooling gives rise to numerous fluctuations to
the topological charge. For improved cooling, plateaus appear after
about forty sweeps and persist for hundreds of sweeps. This is a
celebrated feature of improved cooling.

\begin{figure}[tbp]
\centering{\epsfig{angle=90,figure=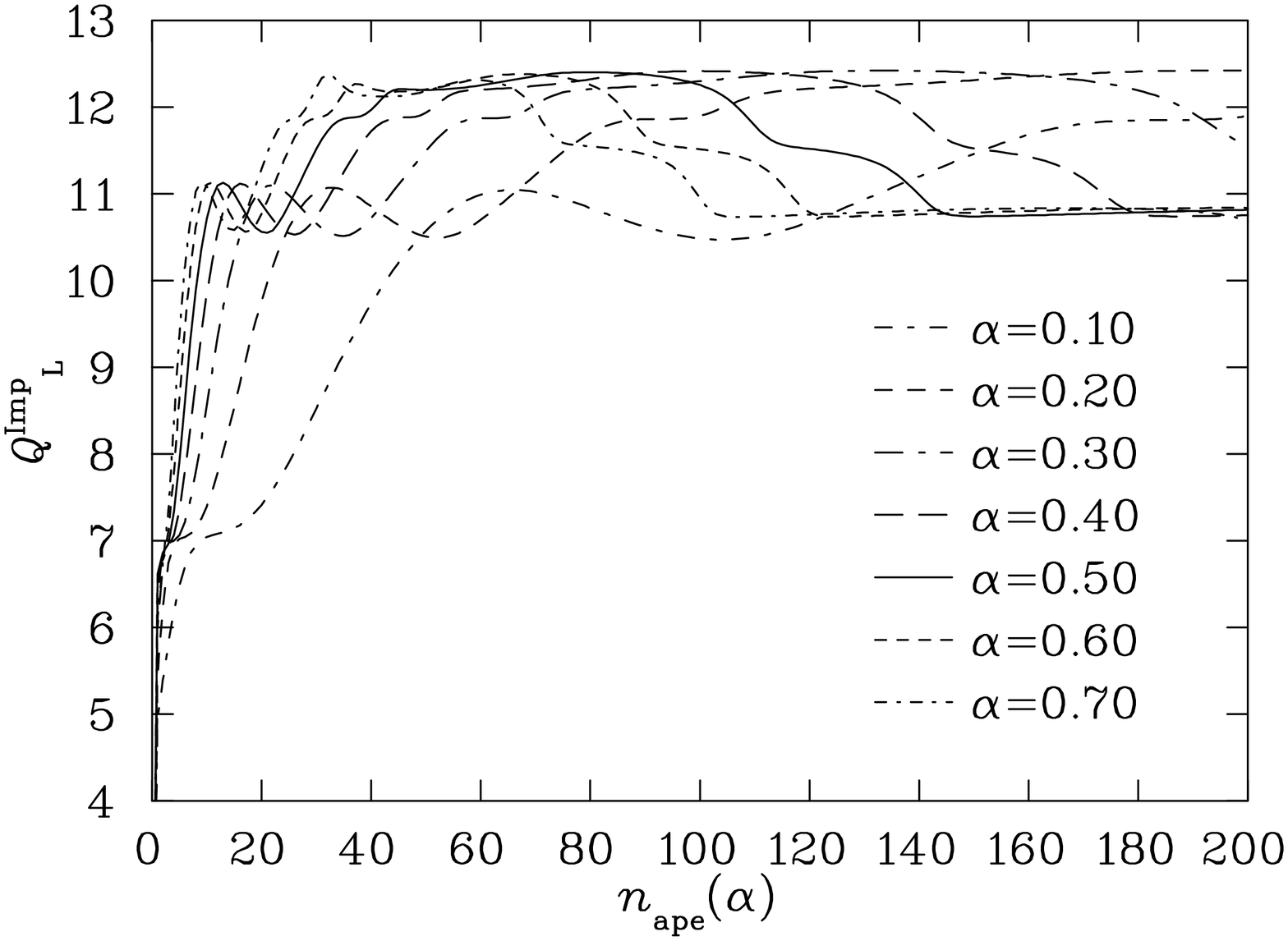,height=9cm} }
\parbox{130mm}{\caption{The evolution of $\qil$ using APE smearing as
a function of APE smearing sweep $\nape$ on the $\1$ lattice at
$\beta=4.38$. Here different line types correspond to different
smearing fractions. }
\label{apeIQc90}}
\end{figure}

\begin{figure}[tbp]
\centering{\epsfig{angle=90,figure=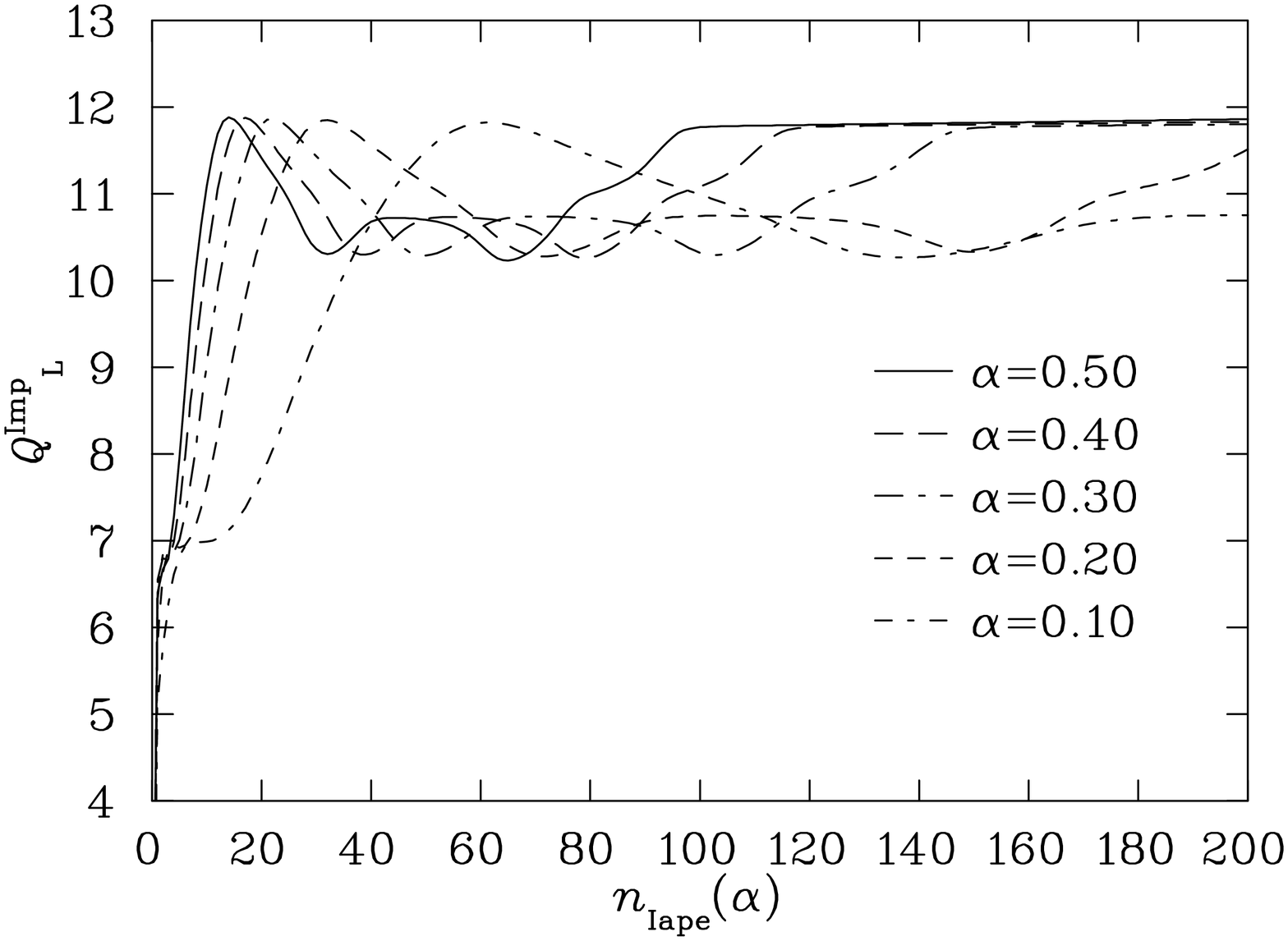,height=9cm} }
\parbox{130mm}{\caption{The evolution of $\qil$ using improved
smearing as a function of APE smearing sweep $\niape$ on the $\1$
lattice at $\beta=4.38$. Here different line types correspond to
different smearing fractions. }
\label{ImpapeIQc90}}
\end{figure}

This algorithmic sensitivity of the topological charge is also seen in
Figs.~\ref{apeIQc90} and~\ref{ImpapeIQc90}, for APE and improved
smearing on a single configuration (the solid line of Figs.~\ref{coolIQc90}
and~\ref{ImpcoolIQc90}). Within APE smearing or improved smearing, the
topological charge trajectories follow similar patterns but different
rates for various smearing fractions. However, APE smearing leads to
values for the topological charge which are different from that
obtained under improved smearing. An important point is that improved smearing
stabilizes the topological charge at 95 sweeps for $\al=0.5$, whereas standard
smearing shows no sign of stability until 140 iterations at $\al=0.5$. Hence we
see significant improvement in the topological aspects of the gauge field configurations
under improved smearing.

\begin{figure}[tbp]
\centering{\epsfig{angle=90,figure=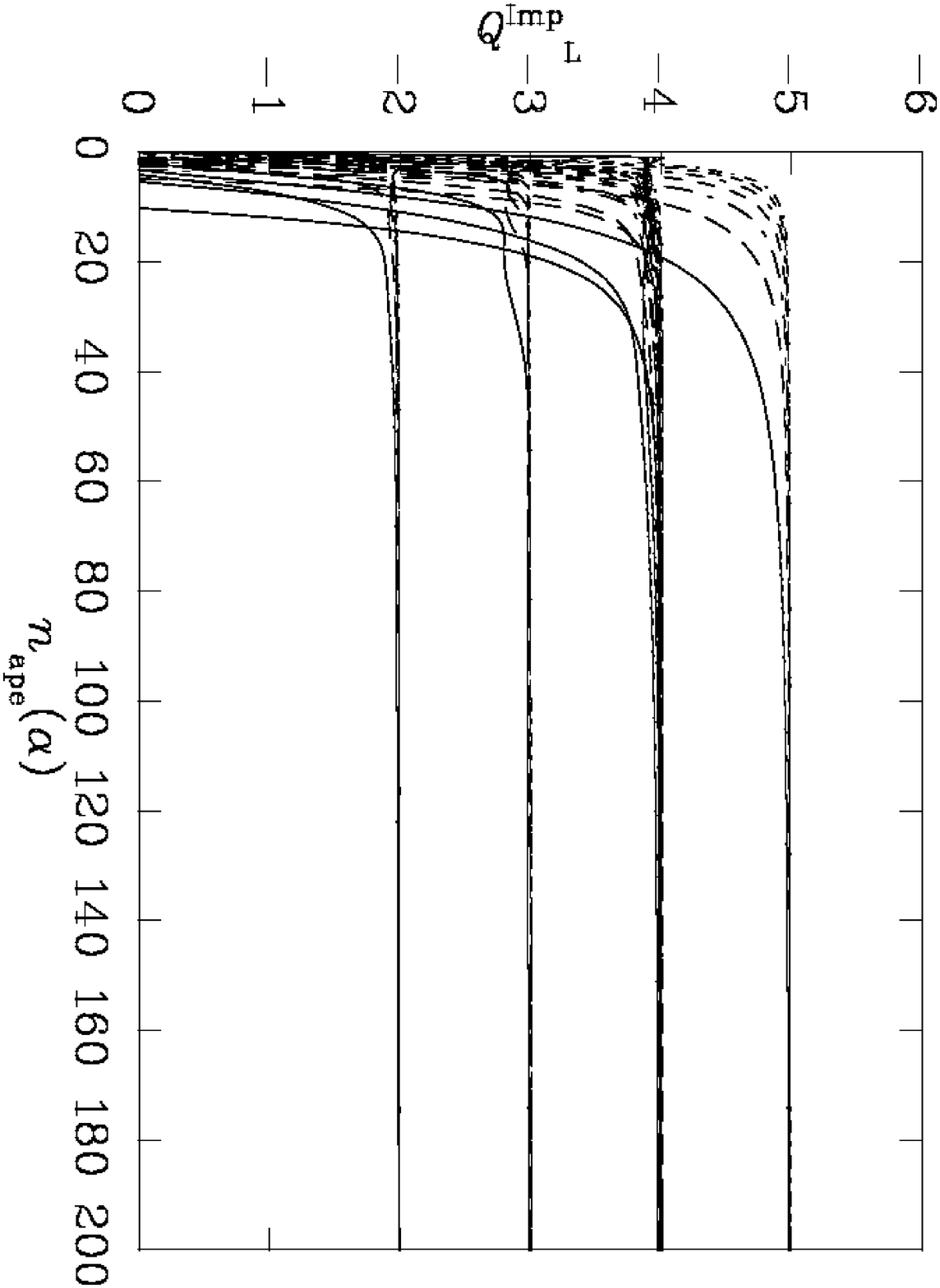,height=9cm} }
\parbox{130mm}{\caption{The evolution of $\qil$ using APE smearing as
a function of APE smearing sweep $\nape$ on the $\2$ lattice at
$\beta=5.00$. Here different line types correspond to different
smearing fractions. }
\label{smearIQc000to005}}
\end{figure}

\begin{figure}[tbp]
\centering{\epsfig{angle=90,figure=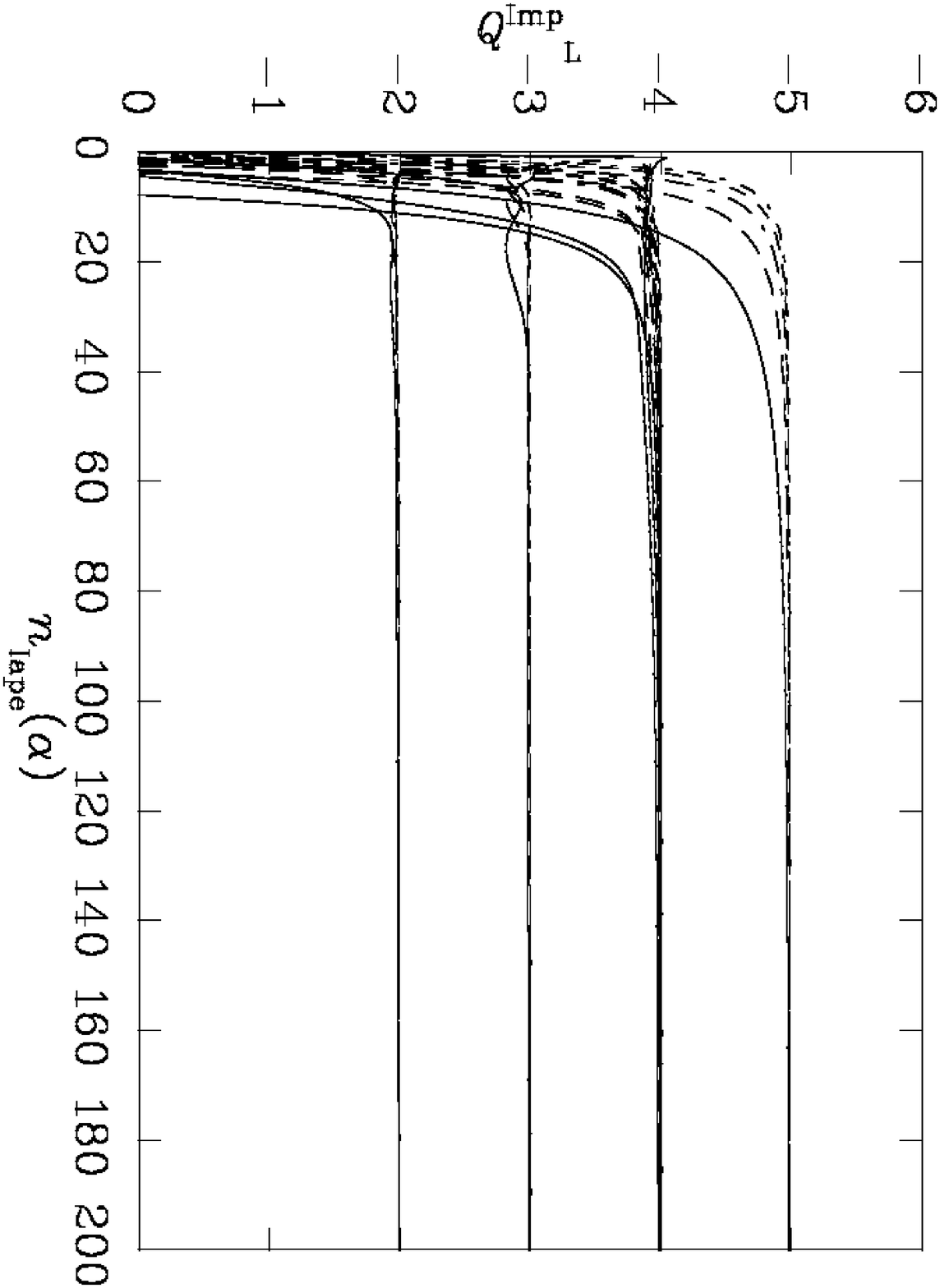,height=9cm} }
\parbox{130mm}{\caption{The evolution of $\qil$ using APE smearing as
a function of APE smearing sweep $\niape$ on the $\2$ lattice at
$\beta=5.00$. Here different line types correspond to different
smearing fractions. }
\label{ImpsmearIQc000to005}}
\end{figure}

On our finer lattice, we find a completely different behavior for the
topological charge evolution. The topological charge is established
very quickly; after a few sweeps in the case of cooling or improved
cooling as illustrated in Figs.~\ref{coolIQc000to005} and
\ref{ImpcoolIQc000to005}. The topological charge persists without
fluctuation for hundreds of sweeps, both for cooling and for smearing
as illustrated in Figs.~\ref{smearIQc000to005} and~\ref{ImpsmearIQc000to005}
for APE and improved smearing respectively. Moreover, the topological
charge is independent of the smearing algorithm.

These results are to be compared with Fig.~\ref{coolIQc90} and
Fig.~\ref{ImpcoolIQc90}, where transitions are observed even with
improved cooling on the coarser lattice with $a = 0.165(2)$ fm. The
results on our fine lattice suggest the characteristic size of
instantons is much larger than the dislocation threshold, such that the
topological structure of the gauge fields is smooth at the scale of
the threshold.

\begin{figure}[tbp]
\centering{\epsfig{angle=90,figure=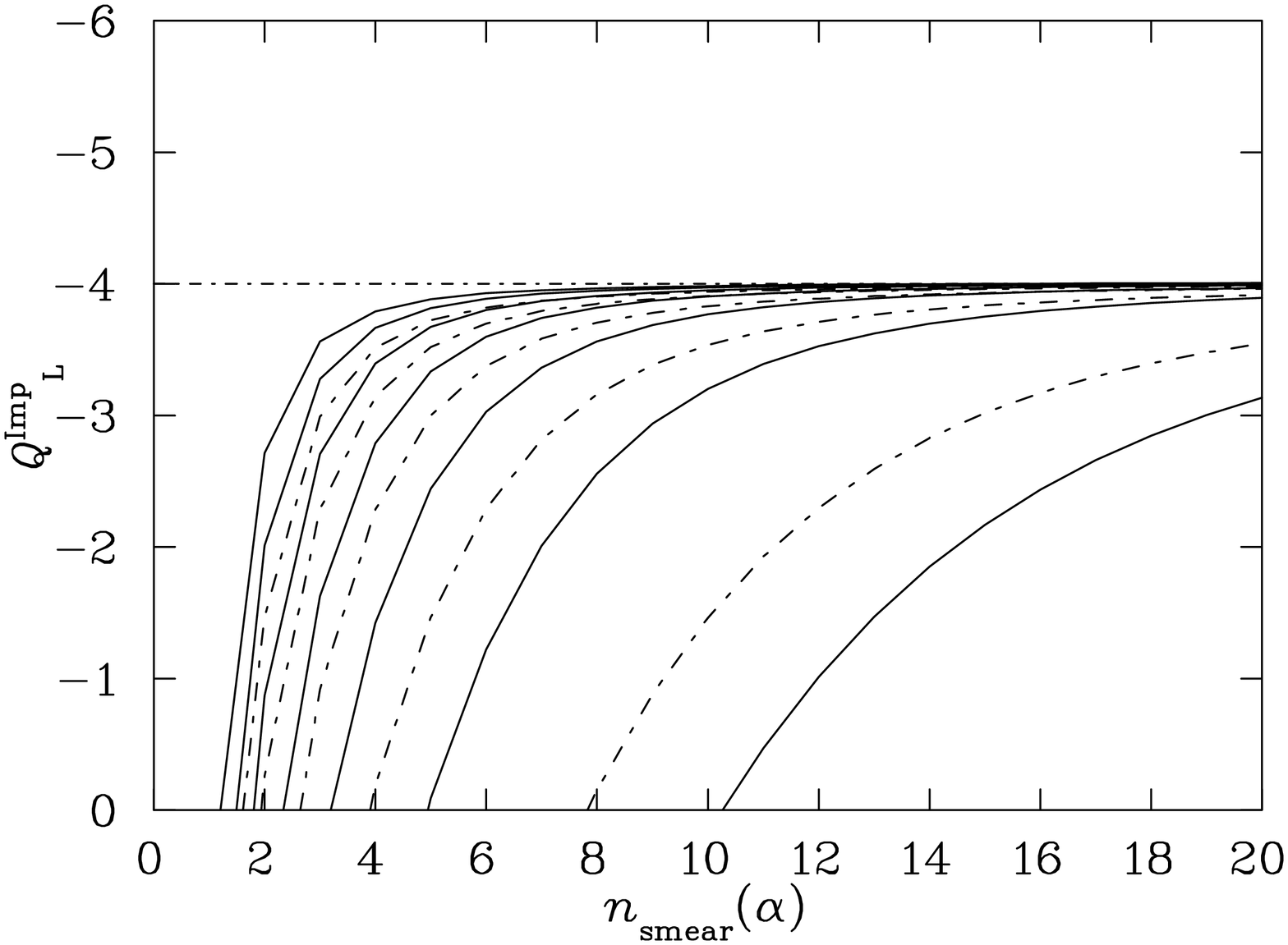,height=9cm} }
\parbox{130mm}{\caption{The evolution of the improved topological
charge, $\qil$, as a function of standard APE smearing sweeps,
$\nape$, for ${0.1}\leq\al\leq{0.7}$ (solid lines) is compared to
improved smearing sweeps, $\niape$, (dotted-dashed lines) for the
same smearing fractions ${0.1}\leq\al\leq{0.5}$ on the $\2$ lattice at
$\beta=5.00$. The horizontal dotted-dashed line is $\qil=-4$.}
\label{bl0to20c002apevsImpape}}
\end{figure}

In Fig.~\ref{smearIQc000to005} for standard APE smearing we observe a
slower convergence to integer topological charge than in Fig.~\ref{ImpsmearIQc000to005}
for improved smearing when ${0.1}\leq\al\leq{0.5}$. This feature of improved
smearing is illustrated in detail in Fig.~\ref{bl0to20c002apevsImpape}. However,
APE smearing has the advantage to allow values for the smearing
fraction up to $\al=0.70$ which cannot be accessed by improved
smearing.

Having demonstrated that it is possible to precisely match the
behavior of the algorithms on fine lattice spacings, we proceed to
calibrate the efficiency of these algorithms in the following
section. 

\section{Smoothing Algorithm Calibration}
\label{calibration}

Here we calibrate the relative rate at which quantum fluctuations are
removed from typical field configurations by the various algorithms.
The calibration is done using the action normalized to the single
instanton action, $S/S_{0}$, on both, the $\1$ lattices and $\2$
lattices. The action normalized to the single instantons action,
$S/S_{0}$, is of particular interest because it provides insight
into the lattice content as well as the rate at which the quantum
fluctuations are removed. 

While there is no doubt that the algorithms may be accurately
calibrated on the fine $\2$ lattice, the $\1$ lattice with $a = 0.165(2)$
fm presents more of a challenge. As a result, in most cases we will
show the graphs produced from the $\1$ lattice analysis and simply
present the numerical results for both the $\1$ and $\2$ lattices.

The numerical results are summarized in Tables \ref{apetab},
\ref{impapetab}, and \ref{cooltab} for the $\1$ lattices and in
Tables~\ref{apetabs24t36}, \ref{impapetabs24t36}, and
\ref{cooltabs24t36} for the $\2$ lattices.

\subsection{APE Smearing and Improved Smearing Calibration.}

To calibrate the rate at which the algorithm reduces the action we
record the nearest number of sweeps required to reach a given
threshold in $S/S_{0}$. The action thresholds are spaced
logarithmically to obtain a uniform distribution in the number of
sweeps required to reach a threshold. The relative rates of smoothing
are established by comparing the relative number of sweeps required to
reach a particular threshold.

\begin{figure}[tbp]
\centering{\epsfig{angle=90,figure=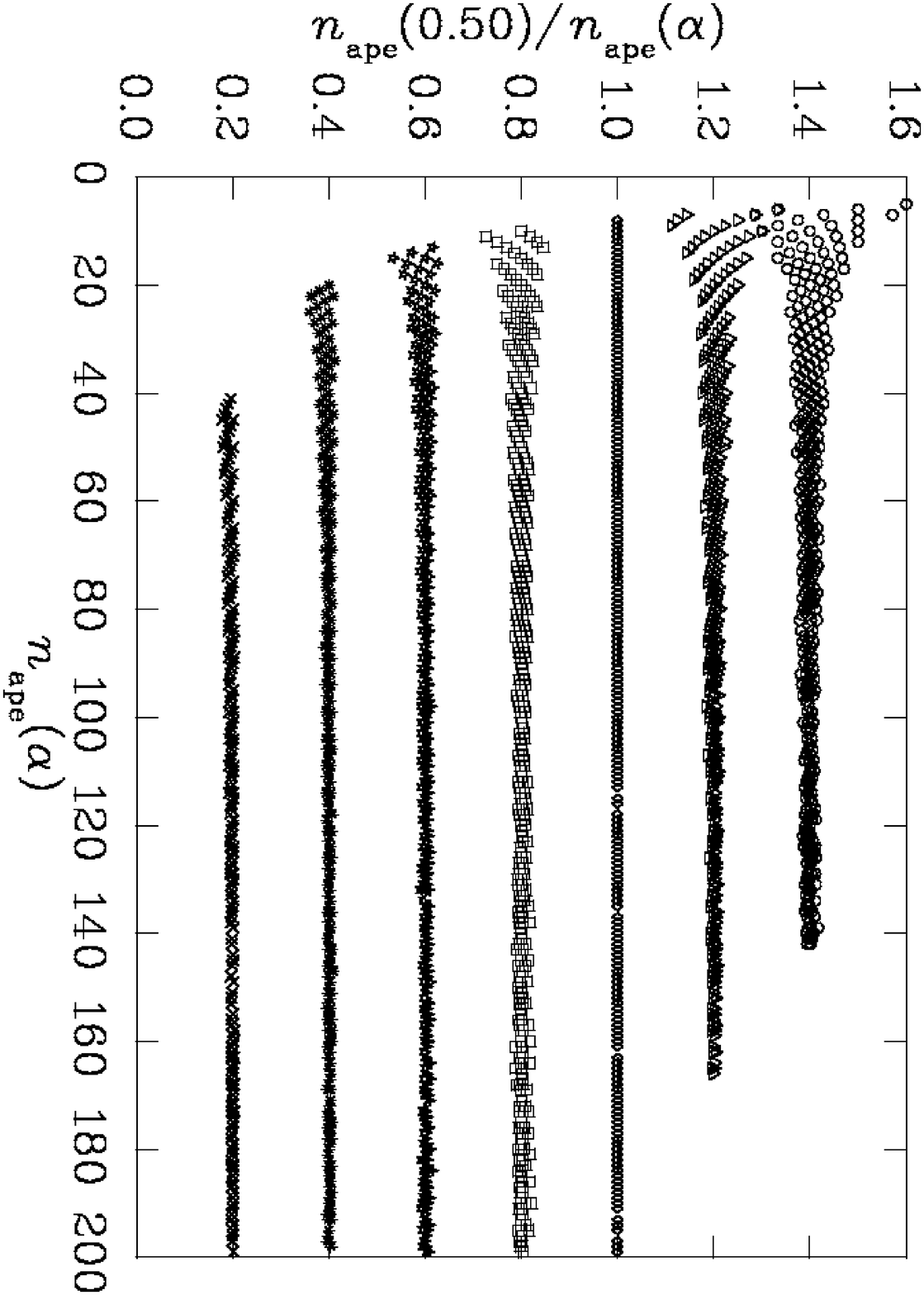,height=9cm} }
\parbox{130mm}{\caption{The ratio $\napefv/\nape$ versus $\nape$ for
numerous $S/S_{0}$ thresholds on the $\1$ lattice at $\bt=4.38$. From
top to bottom the data point bands correspond to $\al=0.7$, 0.6, 0.5,
0.4, 0.3, 0.2, and 0.1. }
\label{ratiosAPE055vsAPE}}
\end{figure}

\begin{figure}[tbp]
\centering{\epsfig{angle=90,figure=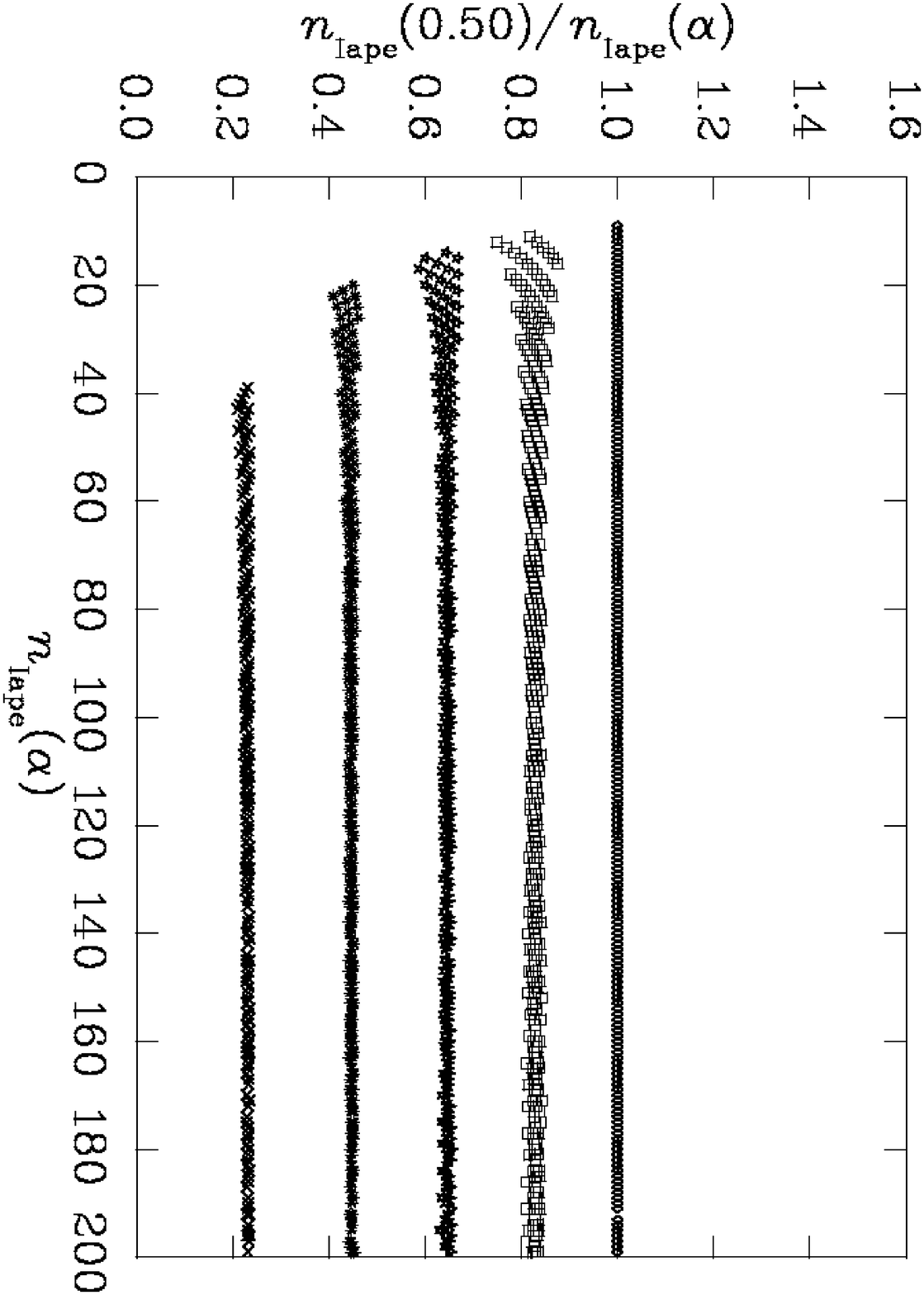,height=9cm} }
\parbox{130mm}{\caption{The ratio $\niapefv/\niape$ versus $\niape$
for numerous $S/S_{0}$ thresholds on the $\1$ lattice at $\bt=4.38$. From
top to bottom the data point bands correspond to
$\al=0.5$, 0.4, 0.3, 0.2, and 0.1.}
\label{ratiosIAPE055vsIAPE}}
\end{figure}

Here we calibrate the APE smearing algorithm characterized by the
smearing fraction $\alpha$ and the number of smearing iterations
$\nape$. The different threshold crossings are characterized by the
number of sweeps required to reach that threshold, $\nape$.
In Fig.~\ref{ratiosAPE055vsAPE} we show the number of sweeps required
to reach a threshold when $\alpha = 0.5$, $\napefv$, relative to
that required for other $\alpha$ values, $\nape$. We plot
these relative smoothing rates as a function of $\nape$ such
that low $S/S_{0}$ thresholds are reached after hundreds of iterations
of the smoothing algorithm. Fig.~\ref{ratiosIAPE055vsIAPE} shows
similar results for improved smearing. In these figures and in
the following analysis, we omit thresholds that result in fewer than
five smoothing iterations as these points produce integer
discretization errors of more than 20\%.

Both standard and improved smearing algorithms have a relative
smoothing rate which is independent of the amount of smoothing
done. By calculating the average value for each of the bands in
Figs.~\ref{ratiosAPE055vsAPE} and~\ref{ratiosIAPE055vsIAPE}, we can
investigate the dependence of the average relative smoothing rate
$\langle \napefv/\nape \rangle$ on $\alpha$.

\begin{figure}[tbp]
\centering{\epsfig{angle=90,figure=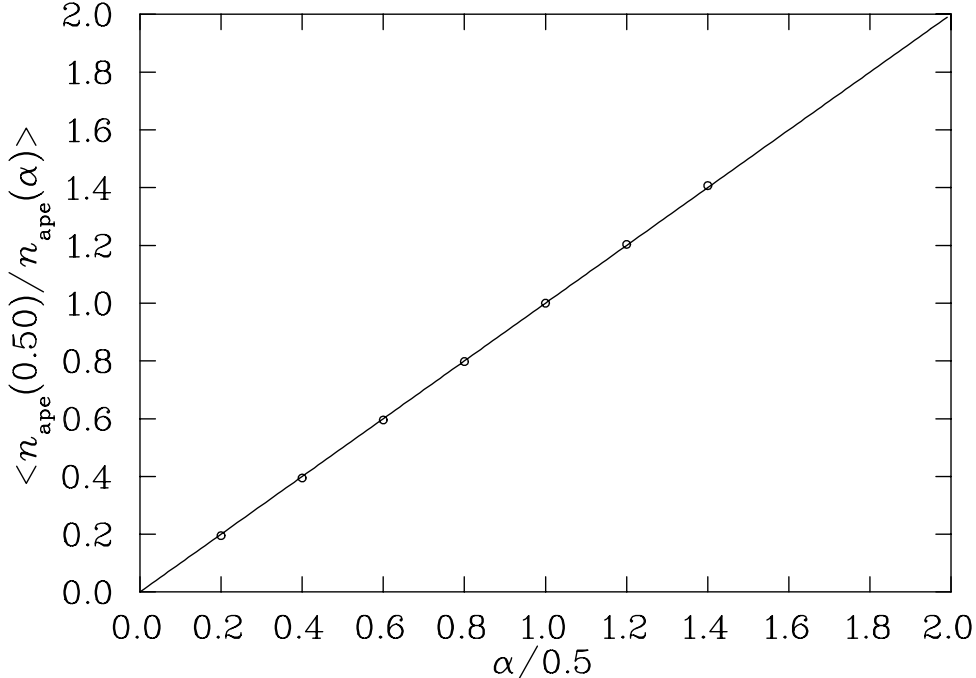,height=9cm} }
\parbox{130mm}{\caption{Illustration of the dependence of
$\langle \napefv/\nape \rangle$ for APE smearing on the smearing
fraction $\alpha$. The solid line is a linear fit to the
data constrained to pass through the origin. }
\label{averagesAPE055vsAPE}}
\end{figure}

Fig.~\ref{averagesAPE055vsAPE} illustrates a linear fit to the data
constrained to pass through the origin. We find 
$\langle \napefv/\nape \rangle = 2.00\, \al$ such that
\begin{equation}
\frac{\napeprm}{\nape} = \frac{\al}{\al^{\prime}}
\label{aperesult}
\end{equation}
in agreement with our earlier analysis \cite{bonnet}. The extent to
which this relationship holds can be verified \cite{bonnet} by
plotting the ratio $\al^{\prime}\napeprm/\al\nape$ and comparing the
results to 1.

\begin{figure}[tbp]
\centering{\epsfig{angle=90,figure=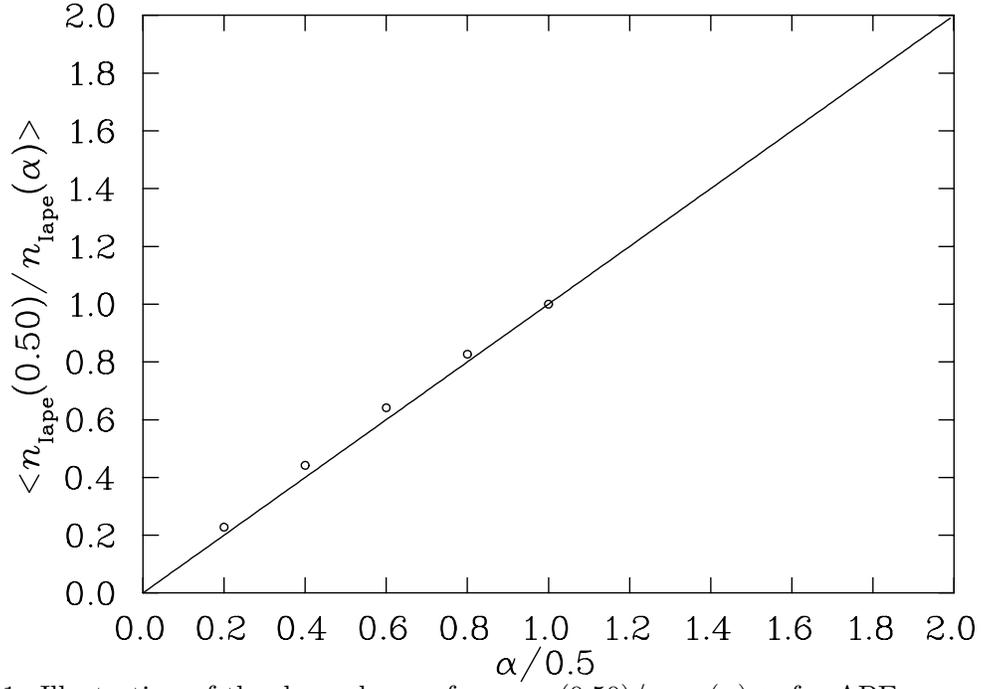,height=9cm} }
\parbox{130mm}{\caption{Illustration of the dependence of
$<\niapefv/\niape>$ for APE smearing on the improved smearing fraction
$\alpha$. The solid line fit is constrained to pass through the origin.}
\label{averagesIAPE055vsIAPE}}
\end{figure}

In plotting the band averages for the improved smearing algorithm
of Fig.~\ref{ratiosIAPE055vsIAPE} in Fig.~\ref{averagesIAPE055vsIAPE},
one finds a small deviation of the points from the line $y=2\, \al$. This
suggests that Eq.~(\ref{aperesult}) is not sufficiently general for
the improved smearing case.

A better approximation to establish the $\al$ dependence, that is
similar to Eq.~(\ref{aperesult}) and contains Eq.~(\ref{aperesult})
is 
\begin{equation}
\frac{\niapeprm}{\niape} = \left(\frac{\al}{\al^{\prime}}\right)^\delta
\, ,
\label{impaperesult}
\end{equation}
where $\delta$ is equal to one in the case of standard
APE smearing and can deviate away from one for improved smearing.

\begin{figure}[tbp]
\centering{\epsfig{angle=90,figure=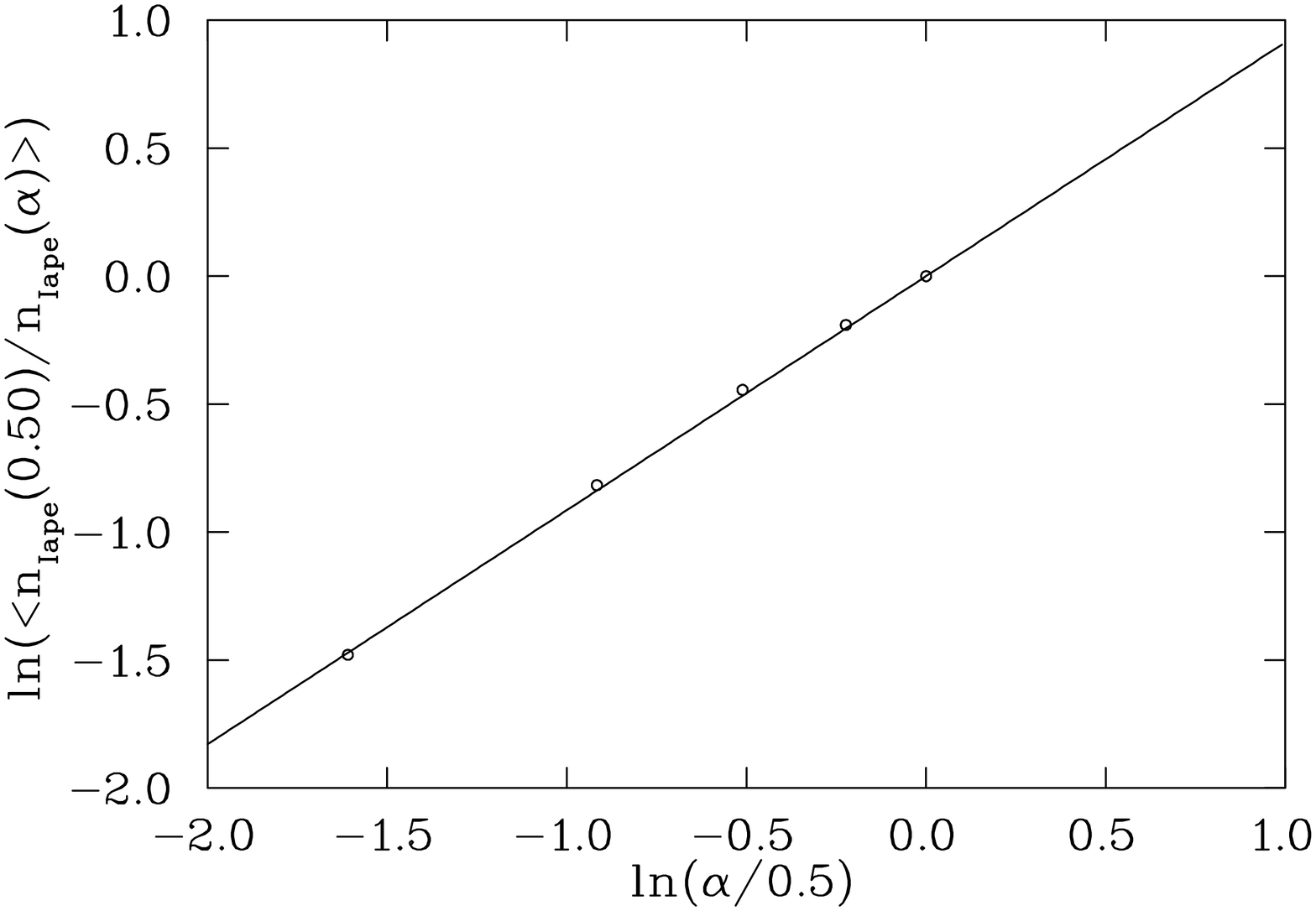,height=9cm} }
\parbox{130mm}{\caption{Illustration of the dependence of
$\ln\left(<\niapefv/\niape>\right)$ on the improved smearing fraction
$\alpha$ for improved smearing. The solid line fit indicates
$\delta = 0.914$.}
\label{btiplot}}
\end{figure}

\begin{figure}[tbp]
\centering{\epsfig{angle=90,figure=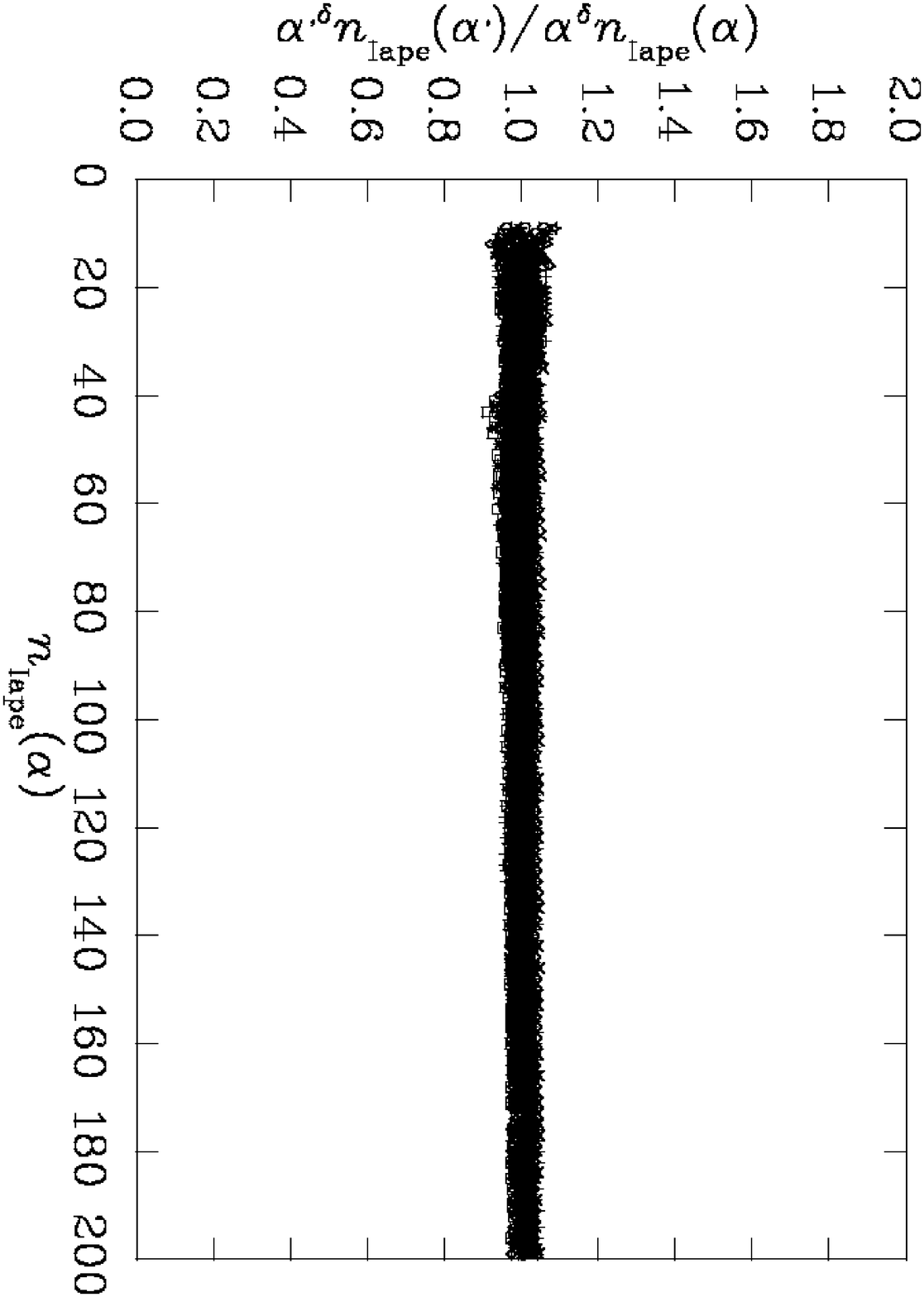,height=9cm} }
\parbox{130mm}{\caption{Illustration of the degree to which the
relation Eq.~(\ref{impaperesult}) is satisfied for improved
smearing. Here the entire data set is plotted for $\al$ and $\alprm=0.5$, 0.4, 0.3, 0.2 and
0.1. Data are from $\1$ lattice at $\bt=4.38$.}
\label{ansatzimpape}}
\end{figure}

In Fig.~\ref{btiplot}, the logarithm of Eq.~(\ref{impaperesult}) is
plotted. The slope of the data provides $\delta = 0.914(1)$ for both
the $\1$ and the $\2$ lattices. We also verified $\delta=1.00$ for the
APE smearing data. Fig.~\ref{ansatzimpape} plots the ratio of the
left- and right-hand sides of Eq.~(\ref{impaperesult}) as a function of
$\niape$ for $\alprm>\al$. The ratio is one as expected with 5\% for
large amounts of smearing where integer discretization errors are
minimized. Throughout the following analysis, $\delta$ is fixed at
0.914(1). 

\subsection{APE and Improved smearing cross calibration.}

\begin{figure}[tbp]
\centering{\epsfig{angle=90,figure=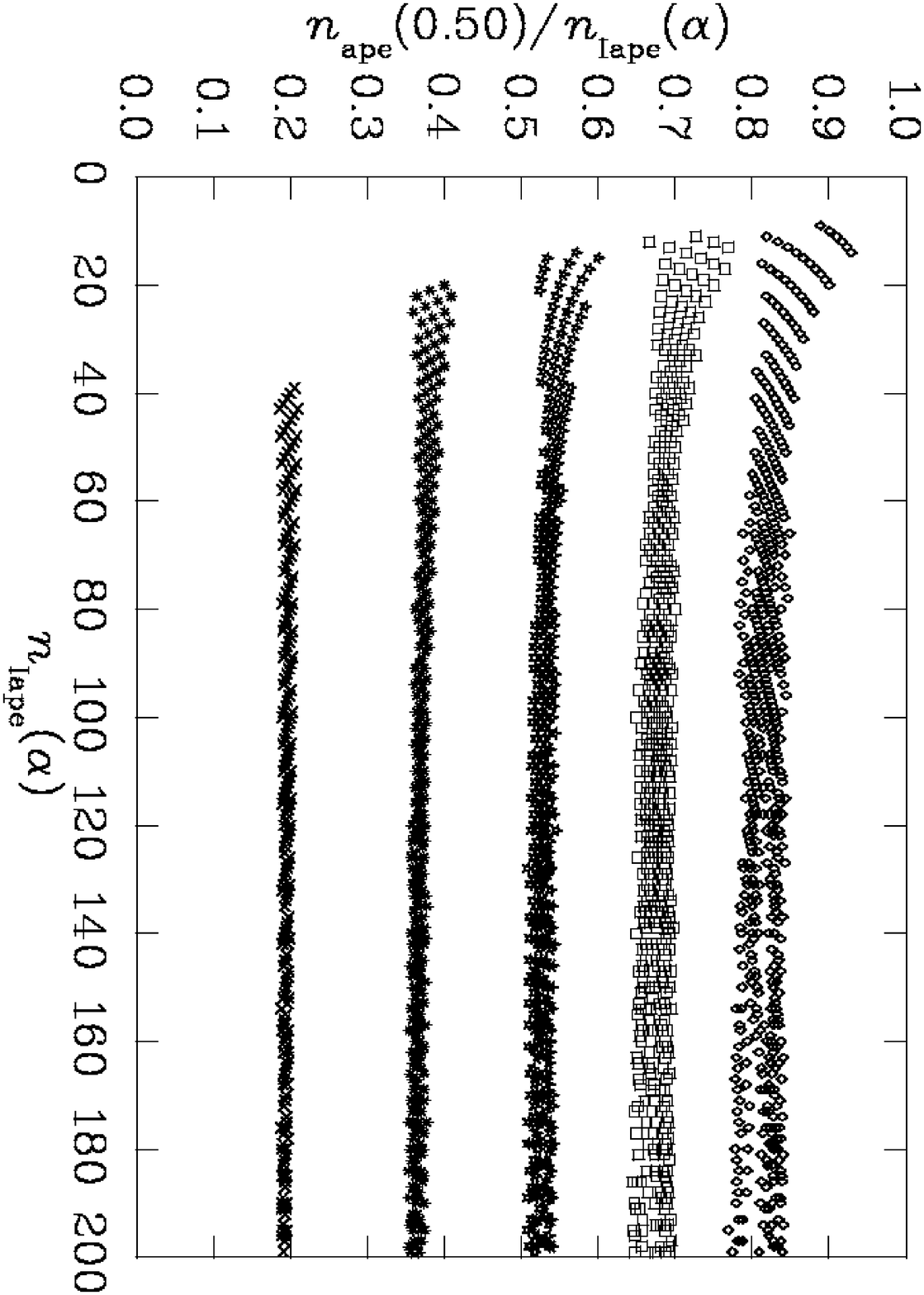,height=9cm} }
\parbox{130mm}{\caption{The ratio $\napefv/\niape$ versus $\niape$ for
numerous threshold actions on the $\1$ lattice at $\bt=4.38$. From top
to bottom the data point bands correspond to improved smearing fractions
$\al=0.5$, 0.4, 0.3, 0.2, and 0.1.}
\label{ratioape055onimpape}}
\end{figure}

In this section we focus on the cross calibration of the smearing
algorithms. In Fig.~\ref{ratioape055onimpape}, we compare the number
of improved smearing sweeps required to reach a threshold for various
improved smearing fractions relative to APE smearing at $\alpha=0.5$.
The lowest band corresponds to an improved smearing fraction of 0.10.
From this, we conclude that for low $\al$ values APE smearing and improved
smearing produce roughly equivalent smeared configurations. However,
there are some evident differences in the rate at which both
algorithms perform. For intermediate to large $\alpha$ there is
curvature in the bands. Early in the smearing process, fewer sweeps
of improved smearing are required to reach a threshold. That is,
improved smearing removes action faster than APE smearing in the
early stages of smearing. This behavior is also manifest in the
analogous results for the fine $\2$ lattice. As emphasized in the
discussion surrounding Fig.~\ref{bl0to20c002apevsImpape}, improved
smearing also provides a topological charge closer to an integer than APE
smearing. Together, these two properties of improved smearing identify
a genuine improvement in the smearing process.

For the coarse $\1$ lattice data, the bands are thick for large
smearing fractions indicating improved smearing does perform
significantly different from standard APE smearing. Contributions
from individual configurations are clearly visible as lines within the
bands. This structure is due to the coarse lattice spacing of
0.165(2) fm which reveals differences between the algorithms. Such
structure is not seen in the fine $\2$ lattice results. There a
precise calibration is possible.

In Tables~\ref{apetab} and \ref{impapetab}, we report the averages of
each band for APE and improved smearing on the $\1$ lattices. In
Tables~\ref{apetabs24t36} and \ref{impapetabs24t36} we report similar
results for the $\2$ lattices.

\begin{table}[tbp]
\caption{The averages of the ratios $<\napefv/\nape>$ and
$<\niapefv/\nape>$ for various smearing fractions $\alpha$ from the
$\1$ lattice at $\bt=4.38$.}
\begin{tabular}{cccccccc}
           &                \multicolumn{7}{c}{$\alpha$ for APE smearing}                                  \\
           & 0.10     & 0.20     & 0.30     & 0.40     & 0.50    & 0.60     & \multicolumn{1}{c}{0.70}     \\
\hline
$\napefv$  & 0.195(1) & 0.394(2) & 0.595(3) & 0.797(4) & 1.0     & 1.203(1) & \multicolumn{1}{c}{1.407(1)} \\
\hline
$\niapefv$ & 0.227(1) & 0.465(1) & 0.706(1) & 0.948(1) & 1.189(1)& 1.431(1) & \multicolumn{1}{c}{1.673(1)}  \\
\hline
\end{tabular}
\label{apetab}
\end{table}

\begin{table}[tbp]
\caption{The averages of the ratios $<\napefv/\niape>$ and
$<\niapefv/\niape>$ for various smearing fractions $\alpha$ from the
$\1$ lattice at $\bt=4.38$.}
\begin{tabular}{cccccc}
           &      \multicolumn{5}{c}{$\alpha$ for improved smearing}               \\
           & 0.10     & 0.20     & 0.30     & 0.40     & \multicolumn{1}{c}{0.50}      \\
\hline
$\napefv$  & 0.196(1) & 0.376(3) & 0.543(1) & 0.697(1) & \multicolumn{1}{c}{0.842(1)}  \\
\hline
$\niapefv$ & 0.228(1) & 0.442(1) & 0.641(1) & 0.827(1) & \multicolumn{1}{c}{1.0}       \\
\hline
\end{tabular}
\label{impapetab}
\end{table}

\begin{table}[tbp]
\caption{The averages of the ratios $<\napefv/\nape>$ and
$<\niapefv/\nape>$ for various smearing fractions $\alpha$ from the
$\2$ lattice at $\bt=5.00$.}
\begin{tabular}{cccccccc}
           &                \multicolumn{7}{c}{$\alpha$ for APE smearing}                                   \\
           & 0.10     & 0.20     & 0.30     & 0.40     & 0.50    & 0.60     & \multicolumn{1}{c}{0.70}      \\
\hline
$\napefv$  & 0.195(1) & 0.395(1) & 0.595(1) & 0.797(1) & 1.0     & 1.205(1) & \multicolumn{1}{c}{1.410(1)}   \\
\hline
$\niapefv$ & 0.227(1) & 0.462(1) & 0.698(1) & 0.937(1) & 1.176(1)& 1.416(1) & \multicolumn{1}{c}{1.658(1)} \\
\hline
\end{tabular}
\label{apetabs24t36}
\end{table}

\begin{table}[tbp]
\caption{The averages of the ratios $<\napefv/\niape>$ and
$<\niapefv/\niape>$ for various smearing fractions $\alpha$ from the
$\2$ lattice at $\bt=5.00$.}
\begin{tabular}{cccccc}
           &      \multicolumn{5}{c}{$\alpha$ for improved smearing}                  \\
           & 0.10     & 0.20     & 0.30     & 0.40     & \multicolumn{1}{c}{0.50}     \\
\hline
$\napefv$  & 0.196(1) & 0.378(1) & 0.546(1) & 0.704(1) & \multicolumn{1}{c}{0.851(1)} \\
\hline
$\niapefv$ & 0.228(1) & 0.442(1) & 0.641(1) & 0.826(1) & \multicolumn{1}{c}{1.0}      \\
\hline
\end{tabular}
\label{impapetabs24t36}
\end{table}

\begin{figure}[tbp]
\centering{\epsfig{angle=90,figure=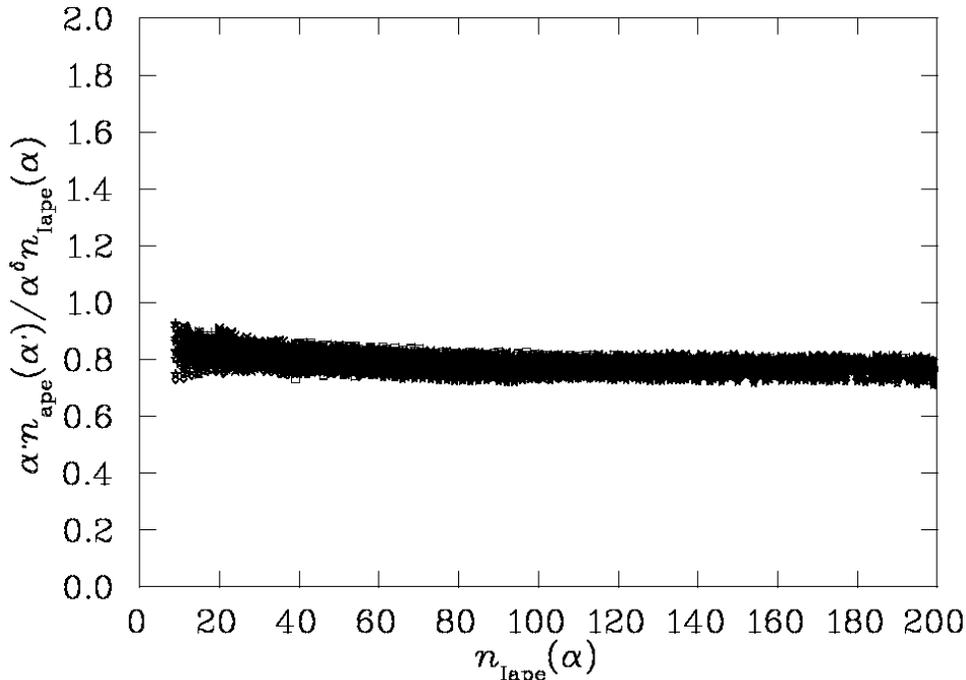,height=9cm} }
\parbox{130mm}{\caption{Illustration of the degree to which the
relation Eq.~(\ref{entirebresult}) is satisfied for calibration of the
action under APE and improved smearing. Here the entire data set
is plotted.}
\label{apevsimpaperule}}
\end{figure}

Based on equations Eq.~(\ref{aperesult}) and Eq.~(\ref{impaperesult}) for APE
and improved smearing, we expect
\begin{equation}
\frac{\alpha' \napeprm}{\alpha^\delta \niape} = {\rm constant} \, .
\label{entirebresult}
\end{equation}
This ratio is plotted in Fig.~\ref{apevsimpaperule} where a rather
mild dependence on $\niape$ is revealed. Averaging these results
provides 0.81(2) for the constant of Eq.~(\ref{entirebresult}). Similar
results are seen for the finer $\2$ lattice, but with greater
precision in the calibration reflected in a narrower band. There the
constant is also 0.81(2). However, it should also be noted that for
$\alpha \le 0.5$, improved smearing achieves integer topological
charge faster than standard APE smearing.

\subsection{Calibration of Cooling and  Smearing}

\begin{figure}[tbp]
\centering{\epsfig{angle=90,figure=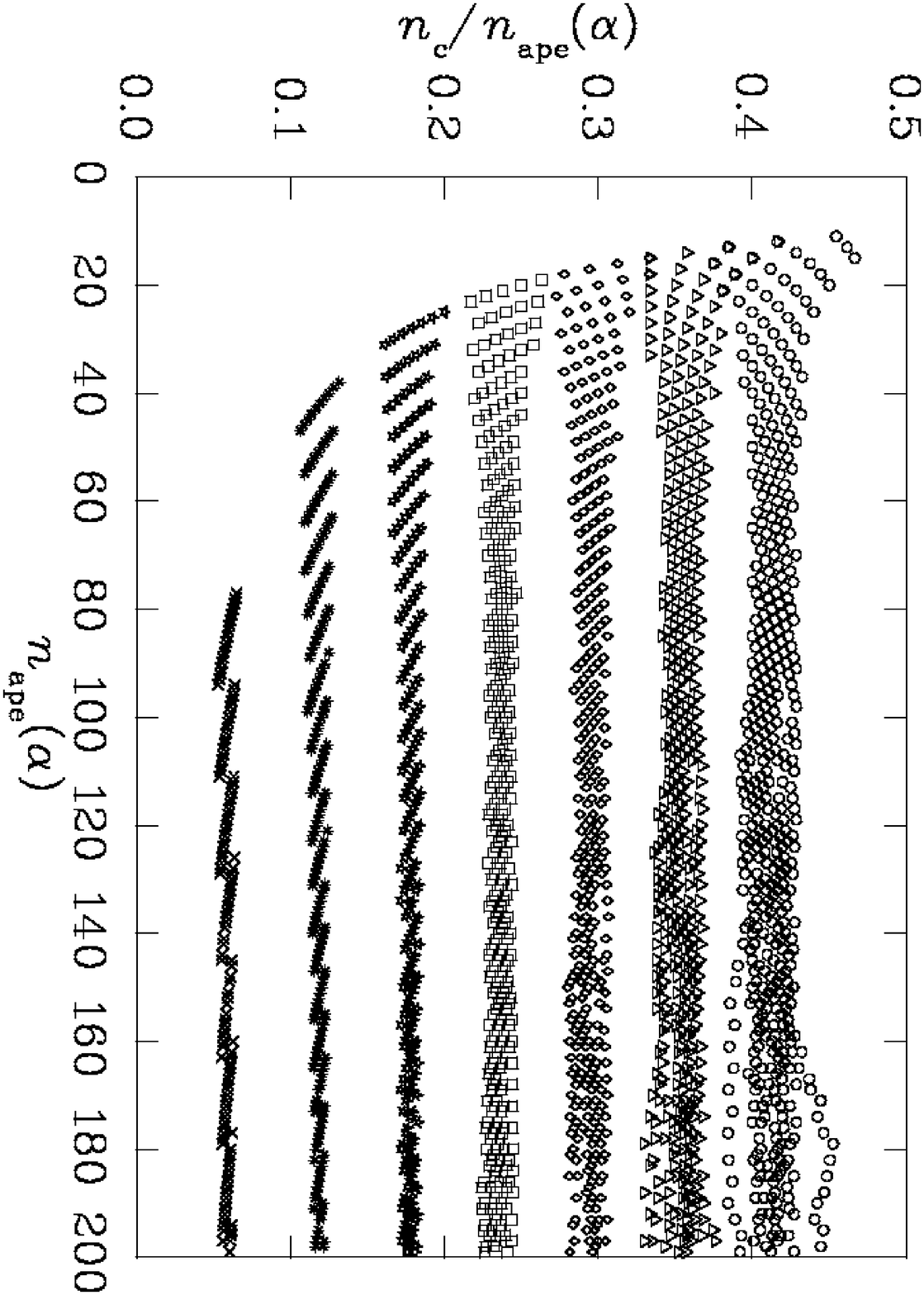,height=9cm} }
\parbox{130mm}{\caption{The ratio $\ncl/\nape$ versus $\nape$ for
numerous action thresholds for the $\1$ lattice at $\bt=4.38$. From top
down the data point bands correspond to $\al=0.7$, 0.6, 0.5, 0.4, 0.3,
0.2, and 0.1.}
\label{coolonape}}
\end{figure}

In this section we apply the ansatz of equations Eq.~(\ref{aperesult}) and
Eq.~(\ref{impaperesult}) to relate the cooling and smearing algorithms.
Fig.~\ref{coolonape} displays results comparing cooling and standard
APE smearing. For $\al$ as small as 0.1 it takes about 75 sweeps of
APE smearing compared with 5 sweeps of cooling to arrive at an
equivalent action. On the other end of the smearing fraction
spectrum, we note the bands become very thick.

The calibration of these ratios indicates
\begin{equation}
\frac{\ncl}{\alpha \nape} = 0.59(1),
\label{ConAPEresult}
\end{equation}
in agreement with that obtained in~\cite{bonnet} where the analysis
was performed on unimproved gauge configurations. The
reduction in ${\cal O}(a^2)$ errors in the gauge field action affect
both algorithms similarly such that the calibration of the relative
smoothing rates remains unaltered.

\begin{figure}[tbp]
\centering{\epsfig{angle=90,figure=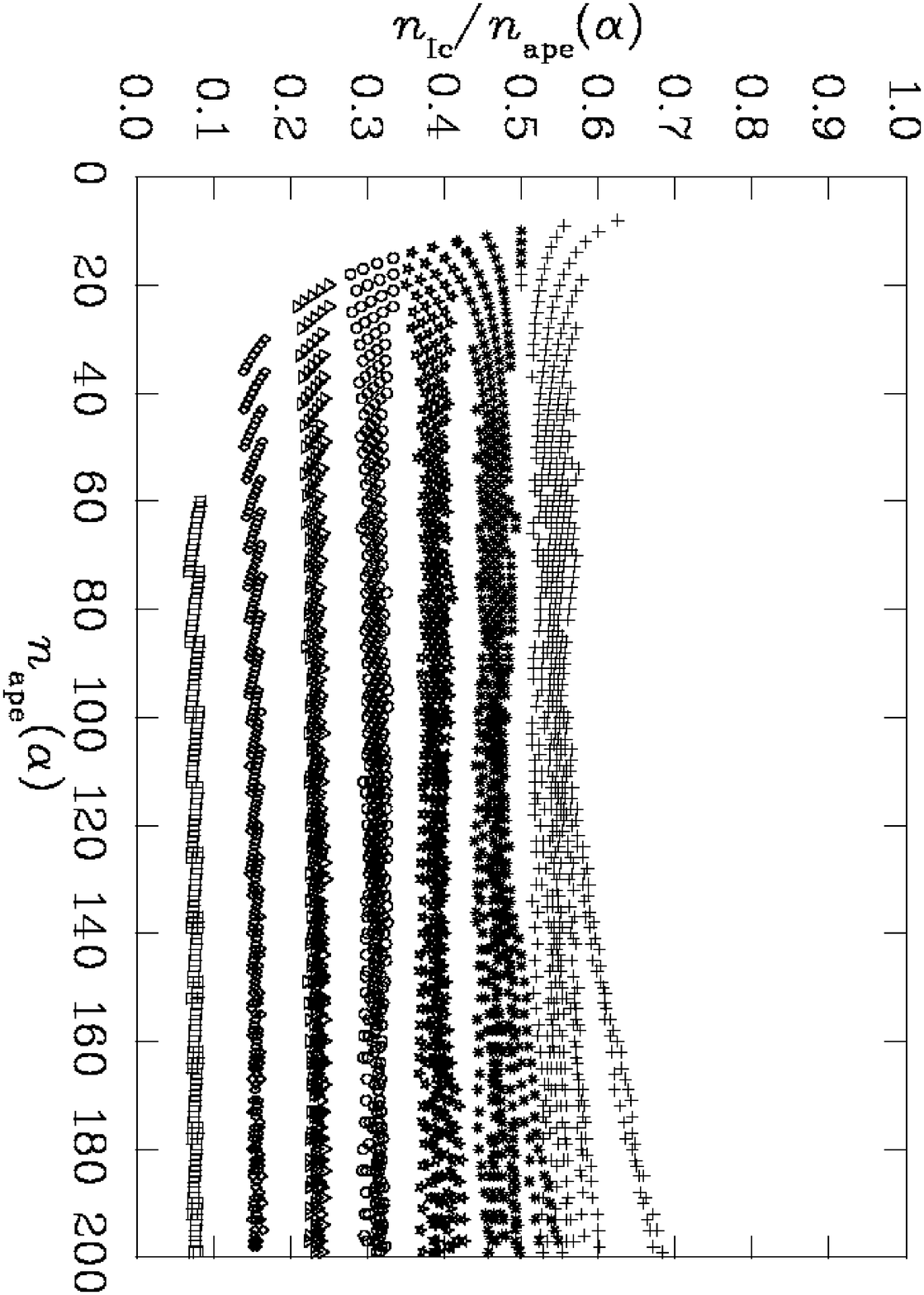,height=9cm} }
\parbox{130mm}{\caption{The ratio $\nicl/\nape$ versus $\nape$ for
numerous action thresholds on the $\1$ lattice at $\bt=4.38$. From top
down the data point bands correspond to $\al=0.7$, 0.6, 0.5, 0.4, 0.3, 0.2,
and 0.1.}
\label{icoolonape}}
\end{figure}

Further broadening of the bands is observed when comparing improved
cooling with APE smearing as illustrated in Fig.~\ref{icoolonape}.
The precision of improved cooling relative to APE smearing leads to
very different smoothed configurations at this coarse lattice spacing
of 0.165(2) fm. This indicates the algorithms are sufficiently
different, that an accurate and meaningful calibration is impossible.

\begin{figure}[tbp]
\centering{\epsfig{angle=90,figure=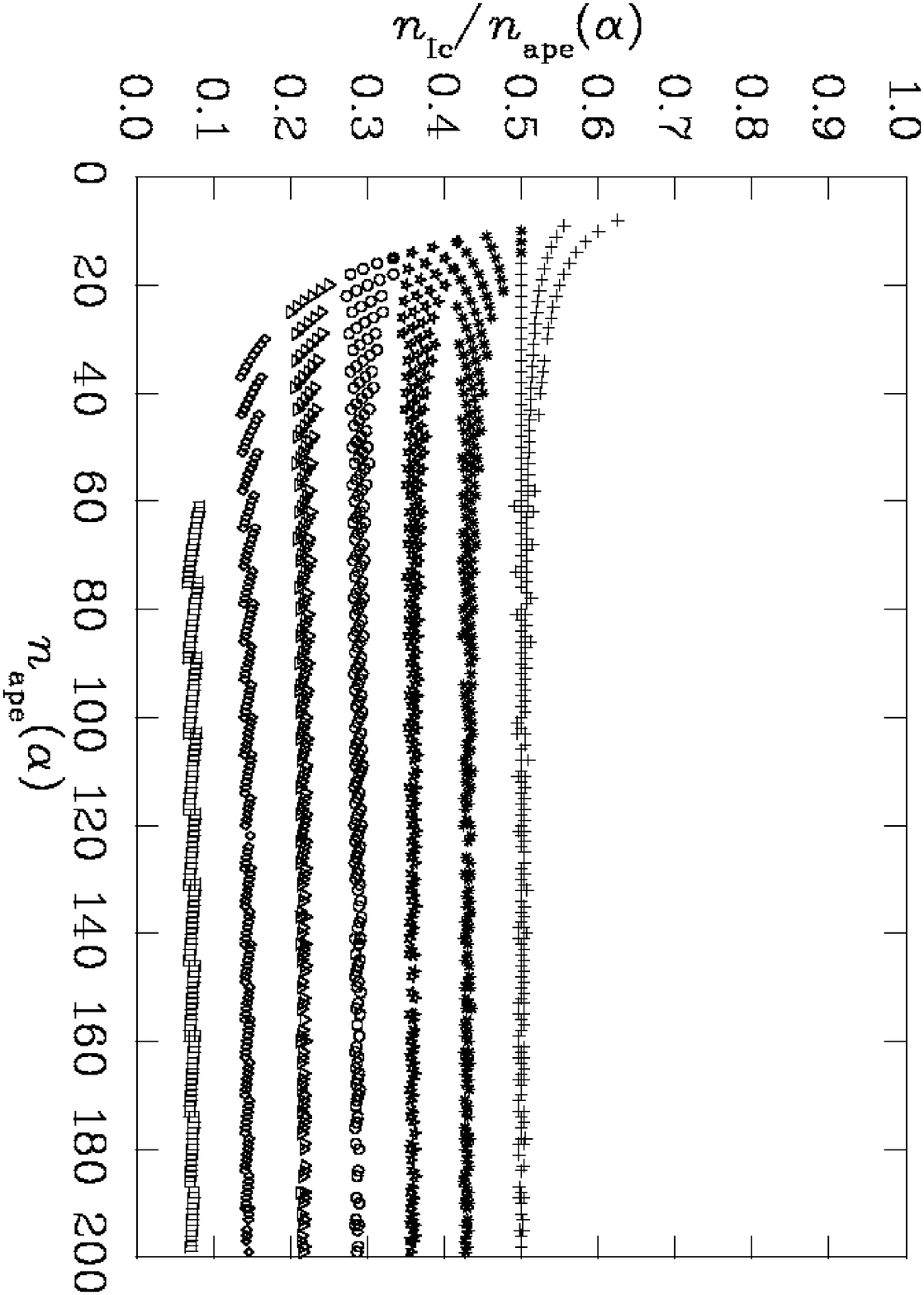,height=9cm} }
\parbox{130mm}{\caption{The ratio $\nicl/\nape$ versus $\nape$ for
numerous action thresholds on the $\2$ lattice at $\bt=5.00$. From top
down the data point bands correspond to
$\al=0.7$, 0.6, 0.5, 0.4, 0.3, 0.2, and 0.1.}
\label{icoolonapes24t36}}
\end{figure}

This effect is not observed when we pass to our fine lattice spacing
as displayed in Fig.~\ref{icoolonapes24t36}. We remind the reader
that the thickness of the band for small numbers of smoothing sweeps
is simply due to the ratio of small integers taken in plotting the
$y$-axis values.

\begin{figure}[tbp]
\centering{\epsfig{angle=90,figure=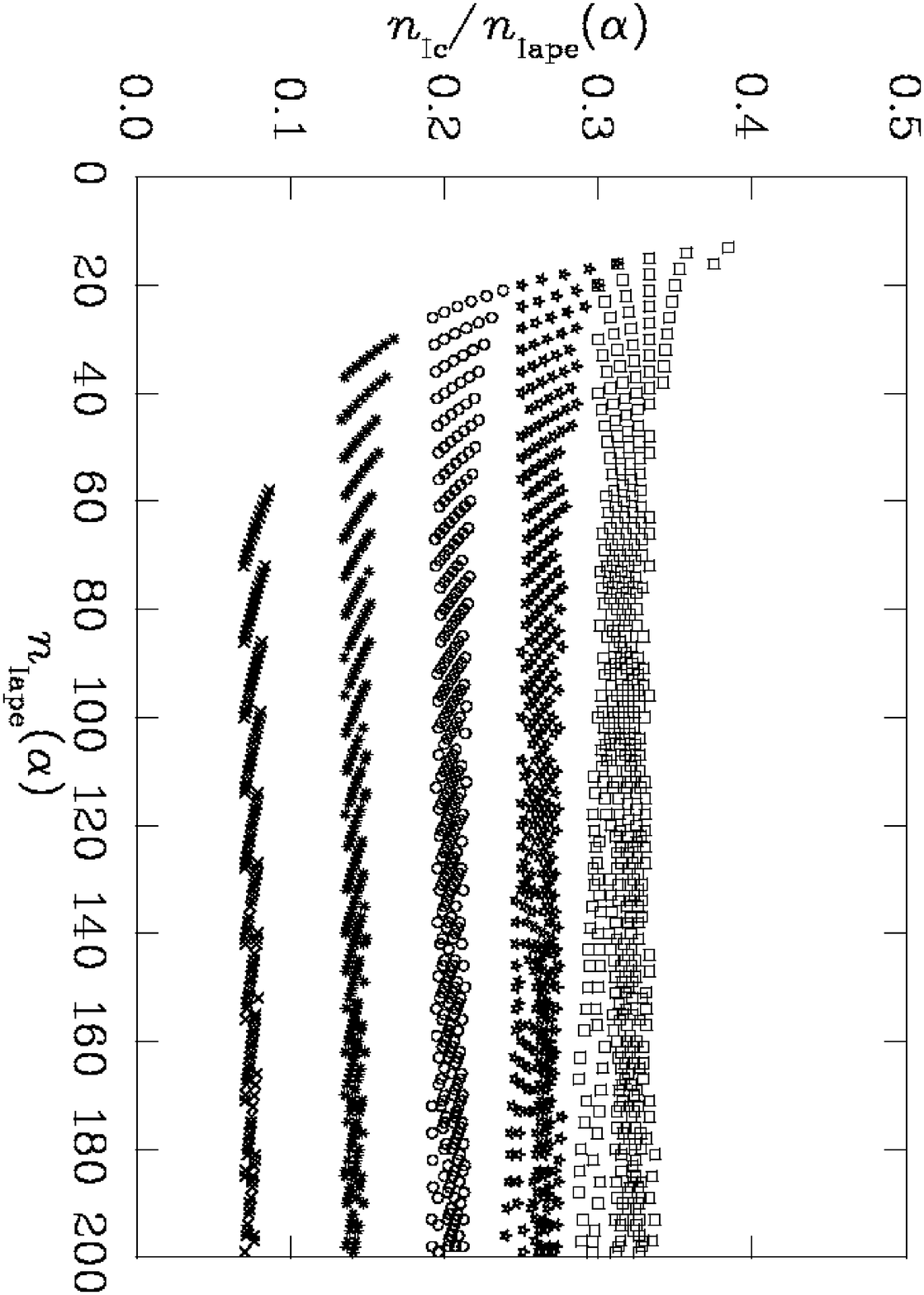,height=9cm} }
\parbox{130mm}{\caption{The ratio $\nicl/\niape$ versus $\niape$ for
numerous action thresholds on the $\1$ lattice at $\bt=4.38$. From
top down the data point bands correspond to $\al=0.5$, 0.4, 0.3, 0.2, and
0.1.}
\label{icoolonimpape}}
\end{figure}

The real test of improved smearing is the extent to which the
algorithm can preserve action associated with topological objects and
thus maintain better agreement with more precise algorithms including
cooling and improved cooling. Fig.~\ref{icoolonimpape} displays
results for the calibration of improved cooling with improved
smearing. Comparing these results for each smearing fraction,
$\alpha$, with that for improved cooling and standard smearing in
Fig.~\ref{icoolonape} reveals that the improved smearing algorithm,
which was seen to be better than standard APE smearing algorithm does not
perform as well as the improved cooling algorithm.

\begin{figure}[tbp]
\centering{\epsfig{angle=90,figure=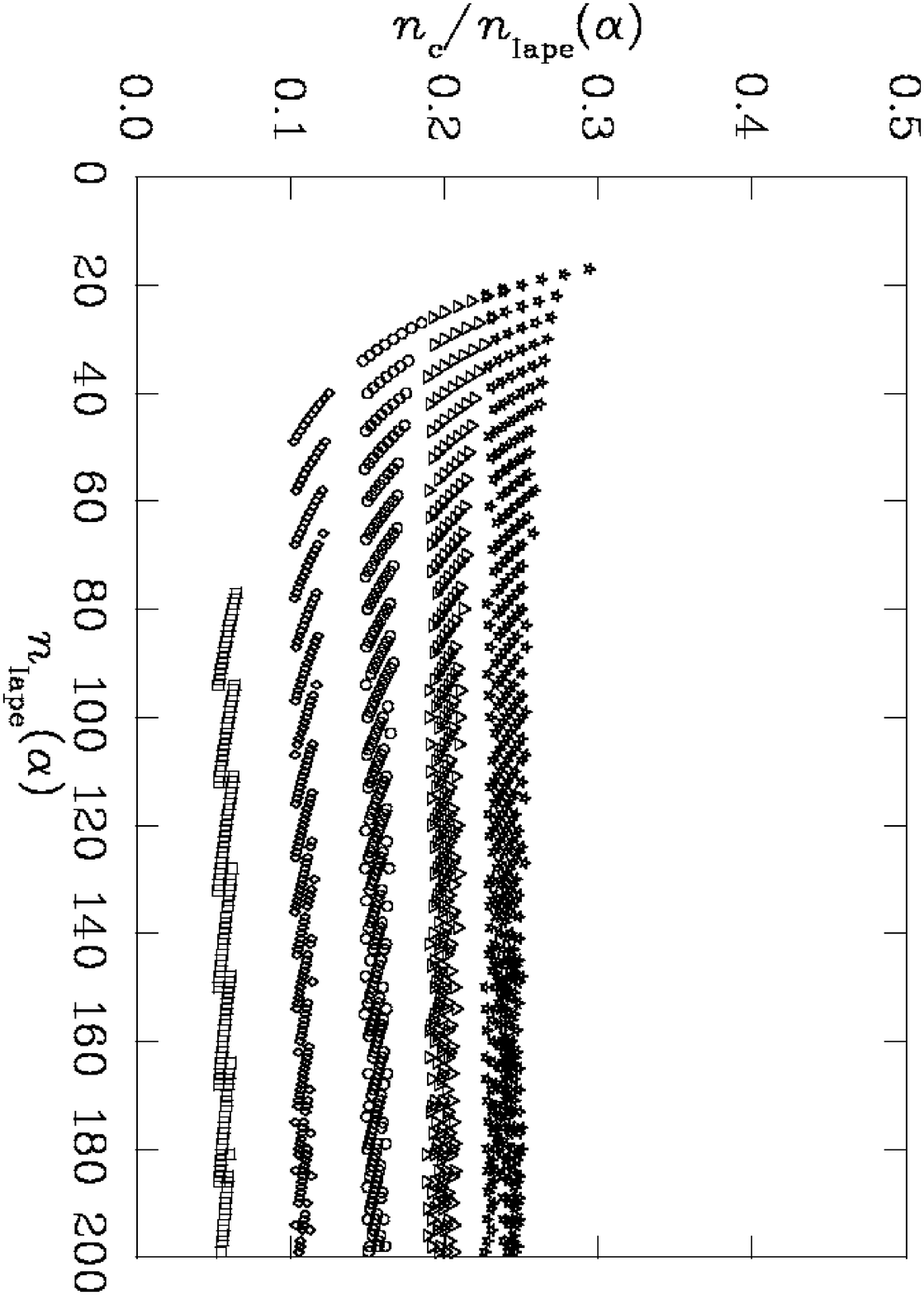,height=9cm} }
\parbox{130mm}{\caption{The ratio $\ncl/\niape$ versus $\niape$ for
numerous action thresholds on the $\1$ lattice at $\bt=4.38$. From
top down the data point bands correspond to $\al=0.5$, 0.4, 0.3, 0.2, and
0.1.}
\label{coolonimpape}}
\end{figure}

Similar results are seen in Fig.~\ref{coolonimpape} where standard cooling is compared
with improved smearing. Hence the annealing of the links in the
process of cooling, where cooled links are immediately passed into the
determination of the next cooled link, is key to the precision with
which cooling can preserve topological structure.

Calibration of the smoothing rates as measured by the action for the
algorithms under investigation are summarized in Tables~\ref{cooltab}
and~\ref{cooltabs24t36}. The entries describe the relative smoothing
rate for the algorithm ratio formed by selecting an entry from the
numerator column and dividing it by the heading of the denominator
columns. The entry comparing APE smearing with itself reports
the level to which the ansatz of equation Eq.~(\ref{aperesult}) is
satisfied. Similarly the entry comparing improved smearing with
itself reports the level to which the ansatz of equation
Eq.~(\ref{impaperesult}) is satisfied.

\begin{table}[tbp]
\caption{Calibration coefficients for various smoothing algorithms on
the $\1$ lattice at $\bt=4.38$. Entries describe the relative
smoothing rate for the algorithm ratio formed by selecting an entry
from the numerator column and dividing it by the heading of the
denominator columns. For example equation Eq.~(\ref{ConAPEresult})
corresponds to the first column of the third row.}
\begin{tabular}{ccccc}
                      &\multicolumn{4}{c}{Denominator} \\
 Numerator            &$\alpha\, \nape$ &$\alpha^\delta \niape$ &$\ncl$  &$\nicl$  \\
\hline		       		  
 $\alprm\, \napeprm$     &1.00(2)          &0.81(2)                &1.69(3) &1.30(2)  \\
 $\alprmdlt\niapeprm$    &1.25(3)          &1.01(2)                &2.13(5) &1.61(3)  \\
 $\ncl$                  &0.59(1)          &0.47(1)                &1       &0.75(1)  \\
 $\nicl$                 &0.77(1)          &0.62(1)                &1.33(2) &1 \\
\end{tabular}
\label{cooltab}
\end{table}

\begin{table}[tbp]
\caption{Calibration coefficients for various smoothing algorithms on
the $\2$ lattice at $\bt=5.00$. Entries describe the relative
smoothing rate for the algorithm ratio formed by selecting an entry
from the numerator column and dividing it by the heading of the
denominator columns. For example equation Eq.~(\ref{entirebresult})
corresponds to the second column of the first row.
}
\begin{tabular}{ccccc}
                      &\multicolumn{4}{c}{Denominator}        \\
 Numerator            &$\alpha\, \nape$ &$\alpha^\delta \niape$  &$\ncl$  &$\nicl$  \\
\hline				        
 $\alprm\, \napeprm$  &1.00(2)          &0.81(2)                 &1.64(2)  &1.37(1)  \\
 $\alprmdlt\niapeprm$ &1.25(3)          &1.01(2)                 &2.04(4)  &1.67(3)  \\
 $\ncl$               &0.611(9)         &0.49(1)                 &1        &0.84(1)  \\
 $\nicl$              &0.734(8)         &0.60(1)                 &1.19(1)  &1        \\
\end{tabular}
\label{cooltabs24t36}
\end{table}

\subsection{Cooling versus Improved cooling.}

\begin{figure}[tbp]
\centering{\epsfig{angle=90,figure=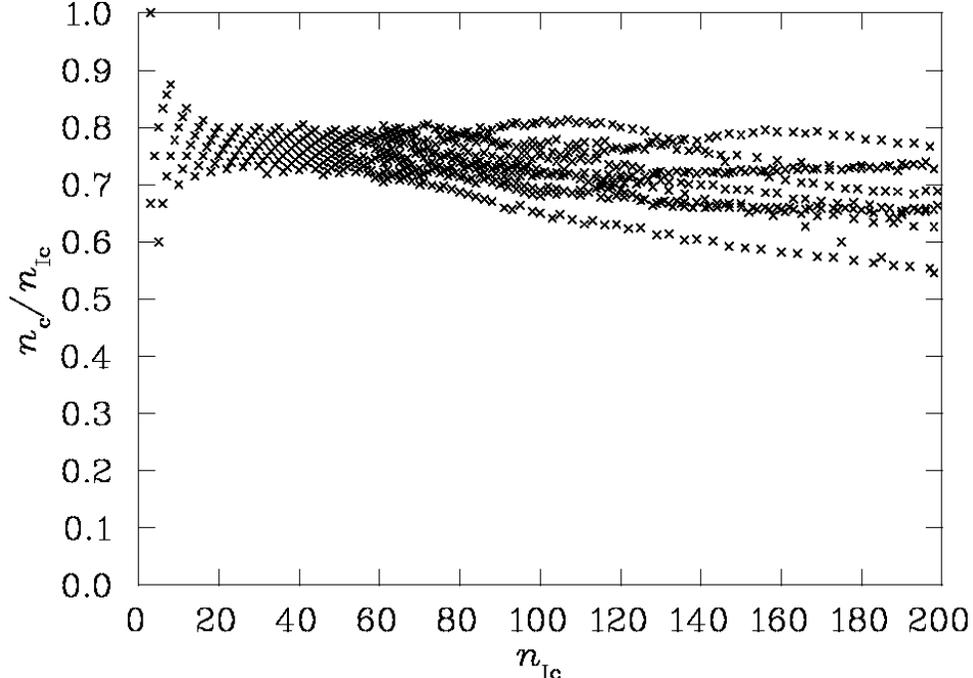,height=9cm} }
\parbox{130mm}{\caption{The ratio $\ncl/\nicl$ versus $\nicl$ for
numerous action thresholds on the $\1$ lattice at $\bt=4.38$. The
significant differences between the algorithms are revealed by the
gauge-configuration dependence of the trajectories.}
\label{coolonicool}}
\end{figure}

\begin{figure}[tbp]
\centering{\epsfig{angle=90,figure=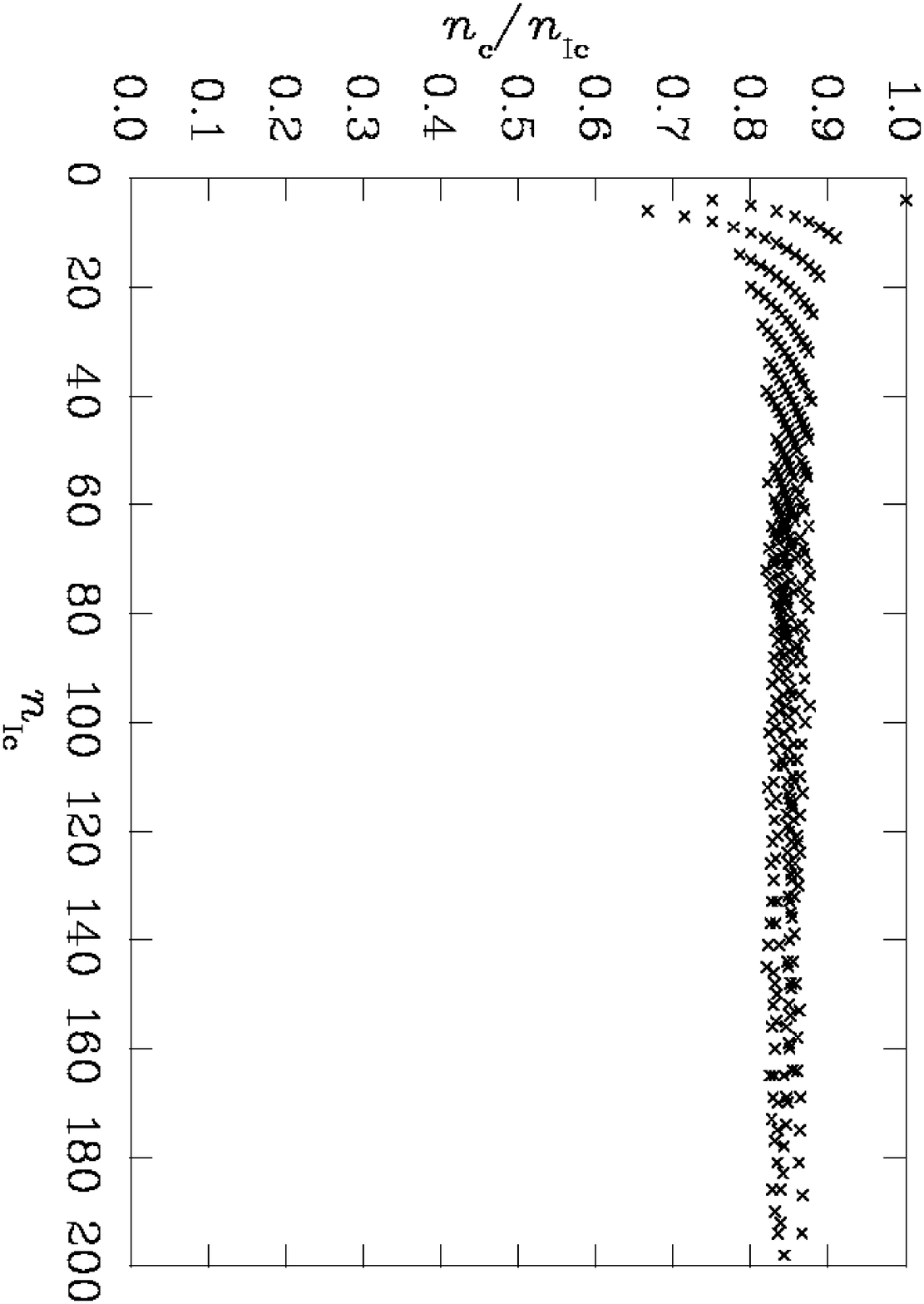,height=9cm} }
\parbox{130mm}{\caption{The ratio $\ncl/\nicl$ versus $\nicl$ for
numerous action thresholds on the $\2$ lattice at $\bt=5.00$.}
\label{coolonicool2}}
\end{figure}

Figure~\ref{coolonicool} reports a comparison of standard cooling with
improved cooling on eleven configurations from the coarse $\1$
lattice. There the ratio $\ncl/\nicl < 1$ confirms the expectation
that standard cooling does not preserve action on the lattice as well
as the ${\cal O}(a^2)$-improved cooling. Fewer standard cooling sweeps are
required to reach the same action threshold. Calibration of the
algorithms appears plausible for the first 80 sweeps of improved
cooling, after which the two algorithms smooth the configurations in
very different manners. Any calibration at this lattice spacing is
only very approximate beyond 80 sweeps of improved cooling where
distinct configuration-dependent trajectories become visible. This
result is contrasted by the analogous analysis on our fine $\2$
lattice illustrated in Fig.~\ref{coolonicool2}. While $\ncl/\nicl$
remains less than one, it is closer to one here than for the coarser
lattice as one might expect.

\section{Conclusions}
\label{conclusion}

We have introduced an improved version of the APE smearing algorithm
founded on the connection between cooling Eq.~(\ref{maxlocal}) and the
projection of the APE smeared link back to the SU(3) gauge group via
Eq.~(\ref{maxAPE1}). This
connection motivates the use of additional extended paths combined
with the standard ``staple'' as governed by the action to reduce the
introduction of ${\cal O}(a^2)$ errors in the smearing projection process.

Clear signs of improvement are observed. For a given smearing
fraction $\alpha$ defined in equation (\ref{apecomb}), improved
smearing preserves the action better than standard APE smearing at each
smearing sweep. At the same time improved smearing brings the improved
topological charge to an integer value faster than standard APE
smearing. 

The extended nature of the ``staple'' in improved smearing reduces the
stability regime for the smearing fraction. We found the improved
smearing algorithm to be stable for $\alpha \leq 0.5$. At $\alpha =
0.6$ the algorithm is unstable whereas standard APE smearing remains
stable for $\alpha \leq 0.75$.

Given the wide variety of smoothing algorithms under investigation in
the field of lattice gauge theory, we have cross calibrated the speed
with which the algorithms remove action from the field configurations.
In particular we have cross calibrated the smoothing rates of APE
smearing at seven values of the smearing fraction; improved smearing
at five values of the smearing fraction; cooling; and improved
cooling. We explored smearing fractions in 0.1 intervals starting at $\alpha =
0.1$.

The calibration has been investigated over a range of 200 sweeps for
each smearing algorithm on ${\cal O}(a^2)$-improved gauge field
configurations. The results of this analysis allows one to make
qualitative comparisons between cooling and smearing algorithms and in
fact make quantitative comparisons of smearing algorithms with
different smearing fractions on lattices as coarse as 0.165(2) fm.
On our fine lattice where the lattice spacing is 0.077(1) fm, the
calibration is quantitative in general.

We have found the relative smoothing rates are described via simple
relationships as reported in Tables \ref{cooltab} and
\ref{cooltabs24t36} for our coarse $\1$ and fine $\2$ lattices
respectively. There the sensitivity of the calibration results on the
lattice spacing may be reviewed. A noteworthy point is that we discovered a necessary correction to the
APE smearing ratio rule \cite{bonnet} when improved smearing is
considered. These algorithms may be calibrated via
\begin{eqnarray}
\frac{\napeprm}{\nape} = \frac{\al}{\alprm} 
\hspace{1cm}{\rm{and}}\hspace{1cm}
\frac{\niapeprm}{\niape} = \left(\frac{\al}{\alprm}\right)^{\delta} \,
\label{ratioruleimp}
\end{eqnarray}
for APE smearing and improved smearing respectively. We find $\delta
= 0.914(1)$ without a significant dependence on the lattice spacing. 

Having cross calibrated these smoothing algorithms, we now proceed to make
contact with physical phenomena~\cite{bernard,perez2,ringwald}. In particular, we note
that it is possible to build in a length scale beyond which cooling does
not affect the links~\cite{perez2}. It would be interesting to explore these
techniques in the context of APE and Improved Smearing. Using a random walk argument,
one can postulate a cooling radius
\bea
r_{\rm{cool}}=c\,\sqrt{\ncl}\,a\,,
\label{radius}
\eea
where $a$ is the lattice spacing and $c$ is a constant independent of $\beta$~\cite{ringwald}.
It has been shown that phenomena taken from simulation results with invariant
$a\sqrt{\ncl}$ scale very well~\cite{ringwald}. The effective range for APE smearing
has been estimated using analytic methods~\cite{bernard}. For small smearing
fraction, $\al$, the effective range is
\bea
r_{\rm{ape}}=\frac{1}{\sqrt{3}}\sqrt{\al\,\nape}\,a.
\label{effrng}
\eea
The product of $\al$ and $\nape$ defines $r_{\rm{ape}}$ in agreement
with the results presented here. Eq.~(\ref{ratioruleimp}) indicates that
this relation holds even for large $\al$. Results of our analysis contained
in tables~\ref{cooltab} and~\ref{cooltabs24t36} allow one to link
Eqs.~(\ref{radius}) and~(\ref{effrng}) and thus determine the constant
$c$. For sufficiently fine lattices $c$ is argued to be independent
of $\bt$~\cite{ringwald} and this is already supported to some
extent by the similarity of the entries in Tables~\ref{cooltab} and~\ref{cooltabs24t36}.
For example, from Table~\ref{cooltabs24t36}, $\ncl={0.611(9)}\,\al\,\nape$ such that
\bea
r_{\rm{cool}}=\frac{1}{\sqrt{3(0.611(9))}}{\sqrt{\ncl}}\,a=0.739(5){\sqrt{\ncl}}\,a\,.\nm
\eea
The effective range for other smoothing algorithms may be derived from Eq.~(\ref{effrng})
in a similarly straight forward manner.

Unfortunately a rigorous analysis of the scaling of the results of Tables~\ref{cooltab} and~\ref{cooltabs24t36}
is not possible. We have clear evidence that the topology of Yang-Mills gauge fields
cannot be reliably studied using the algorithms presented here on lattice spacings as coarse as 0.165(2) fm.
Different algorithms lead to different topological charges, differing
quite widely in some cases as reported in Figs.~\ref{coolIQc90}
and~\ref{ImpcoolIQc90}. Moreover, subtle differences in the
cooling algorithms can lead to different topological charge
determinations as illustrated in Figs.~\ref{nsubstd} and~\ref{nsubimp}.
As discussed in Sec.~\ref{subgroups}, the proximity of the dislocation
thresholds of the algorithms to the typical size of instantons and variations
in the threshold from one algorithm to another causes some (anti)instantons to survive
under one algorithm, whereas they are removed under another.

In contrast, the fine $\2$ lattice results, where $a = 0.077(1)$ fm,
display excellent agreement among every smoothing algorithm
considered. In this case it appears that the dislocation thresholds are
smaller than the characteristic size of topological fluctuations\footnote{We define a ``topological fluctuation''
to refer to objects with $Q=\pm{1}$ but $S/\szero>1$.} such that the gauge
fields are already sufficiently smooth to unambiguously extract the
topology of the gauge fields.

\begin{figure}[tbp]
\centering{\epsfig{angle=0,figure=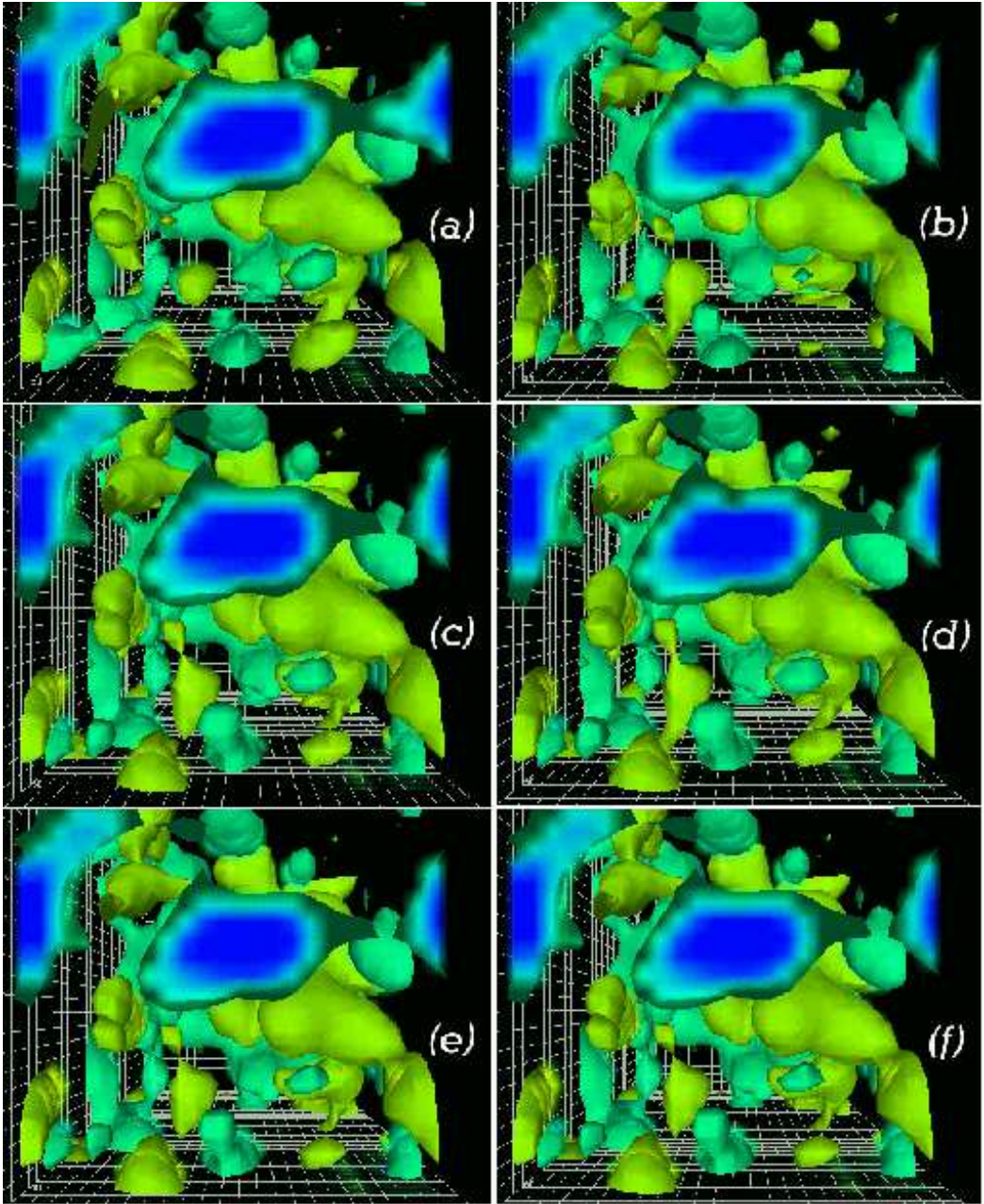,height=17cm} }
\parbox{130mm}{\caption{The topological charge density of a $\2$ lattice for
fixed $x$ coordinate. The instantons (anti-instantons) are colored
red to yellow (blue to green). Fig.~a) shows the topological charge
density after 9 cooling sweeps. Each of the following figures display
the result of a different smoothing algorithm calibrated according to
Table \protect\ref{cooltabs24t36} to reproduce as closely as possible
the results depicted in Fig.~a). Fig.~b) illustrates the
topological charge density after 11 sweeps of improved cooling. Fig.~\
c) shows the topological charge density after 21 APE smearing steps at
$\al=0.70$. Fig.~d) illustrates the topological charge density after
49 APE smearing steps at $\al=0.30$. In Fig.~e) the topological
charge density is displayed after 35 sweeps of improved smearing at
$\al=0.50$. Finally, Fig.~f) shows the topological charge density
after 55 sweeps of improved smearing at $\al=0.30$. Apart from Fig.~b)
for improved cooling, which differs largely due to round off in the
sweep number, all the plots compare very favorably with each other.}
\label{Visual}}
\end{figure}

As a final comparison of the smoothing algorithms, we provide a visual
representation of a gauge field configuration after applying various
smoothing algorithms. Figure~\ref{Visual} illustrates a rendering of
the topological charge density for a slice of one of the fine $\2$
lattice configurations. While our calibration has been carried out by
considering the total action of the gauge fields, the following
analysis allows us to examine the extent to which the calibration is
accurate at a microscopic level.

In Fig.~\ref{Visual}, red shading indicates large positive
topological charge density with decreasing density becoming yellow in
color, while blue shading indicates large in magnitude, negative
topological charge density decreasing in magnitude through the color
green. Here cooling (a), improved cooling (b), APE smearing at
$\al=0.70$ (c), APE smearing at $\al=0.30$ (d), improved smearing at
$\al=0.50$ (e) and improved smearing at $\al=0.30$ (f), are compared
at the number of smoothing iterations required for each algorithm to
produce an approximately equivalent smoothed gauge field
configuration. While Fig.~b) for improved cooling differs somewhat
due to round off in the sweep number, the remaining plots compare very
favorably with each other. These visualizations confirm that the
different smoothing algorithms considered in this investigation can be
accurately related via the calibration analysis presented here and
summarized in Tables \ref{cooltab} and \ref{cooltabs24t36}.

\section*{Acknowledgment}

Thanks to Francis Vaughan of the South Australian Centre for Parallel
Computing and the Distributed High-Performance Computing Group for for
generous allocations of time on the University of Adelaide's CM-5.
Support for this research from the Australian Research Council is
gratefully acknowledged.

\end{document}